\documentclass[conference, 9pt]{IEEEtran}
\IEEEoverridecommandlockouts
\usepackage{cite}
\usepackage{amsmath,amssymb,amsfonts}
\usepackage{algorithmic}
\usepackage{graphicx}
\usepackage{textcomp}
\usepackage{array}
\usepackage[table,xcdraw]{xcolor}
\usepackage{booktabs}
\usepackage{algorithm}

\usepackage{subfig}
\usepackage{lineno,hyperref}

\makeatletter
\DeclareRobustCommand\onedot{\futurelet\@let@token\@onedot}
\def\@onedot{\ifx\@let@token.\else.\null\fi\xspace}

\makeatother

\newcommand{\tref}[1]{Tab.~\ref{#1}}

\newcommand{\fref}[1]{Fig.~\ref{#1}}

\newcommand{\sref}[1]{Sec.~\ref{#1}}

\newcommand{\aref}[1]{Algo.~\ref{#1}}

\def\BibTeX{{\rm B\kern-.05em{\sc i\kern-.025em b}\kern-.08em
    T\kern-.1667em\lower.7ex\hbox{E}\kern-.125emX}}
\begin{document}

\newcommand{\method}{LS-Gaussian}

\title{No Redundancy, No Stall: Lightweight Streaming 3D Gaussian Splatting for Real-time Rendering}

\author{\IEEEauthorblockN{Linye Wei$^{1,2}$, Jiajun Tang$^{3,4}$, Fan Fei$^{3,4}$, Boxin Shi$^{3,4}$, Runsheng Wang$^{2,5,6}$, Meng Li$^{1,2,6*}$}
\IEEEauthorblockA{$^1$Institute for Artificial Intelligence, Peking University, Beijing, China}
\IEEEauthorblockA{$^2$School of Integrated Circuits, Peking University, Beijing, China}
\IEEEauthorblockA{$^3$ State Key Laboratory of Multimedia Information Processing, School of Computer Science, Peking University}
\IEEEauthorblockA{$^4$National Engineering Research Center of Visual Technology, School of Computer Science, Peking University}
\IEEEauthorblockA{$^5$Institute of Electronic Design Automation, Peking University, Wuxi, China}
\IEEEauthorblockA{$^6$Beijing Advanced Innovation Center for Integrated Circuits, Beijing, China}
\IEEEauthorblockA{meng.li@pku.edu.cn}
\thanks{This work was supported in part by NSFC under Grant 62495102, Grant 92464104, Grant 62125401, and Grant 62136001, in part by Beijing Outstanding Young Scientist Program under Grant JWZQ20240101004, in part by Beijing Municipal Science and Technology Program under Grant Z241100004224015, and in part by 111 Project under Grant B18001.}}

\maketitle

\begin{abstract}

3D Gaussian Splatting (3DGS) enables high-quality rendering of 3D scenes and is getting increasing adoption in domains like autonomous driving and embodied intelligence. However, 3DGS still faces major efficiency challenges when faced with high frame rate requirements and resource-constrained edge deployment. To enable efficient 3DGS, in this paper, we propose \method, an algorithm/hardware co-design framework for lightweight streaming 3D rendering. \method~is motivated by the core observation that 3DGS suffers from substantial computation redundancy and stalls. On one hand, in practical scenarios, high-frame-rate 3DGS is often applied in settings where a camera observes and renders the same scene continuously but from slightly different viewpoints. Therefore, instead of rendering each frame separately, \method~proposes a viewpoint transformation algorithm that leverages inter-frame continuity for efficient sparse rendering. On the other hand, as different tiles within an image are rendered in parallel but have imbalanced workloads, frequent hardware stalls also slow down the rendering process. \method~predicts the workload for each tile based on viewpoint transformation to enable more balanced parallel computation and co-designs a customized 3DGS accelerator to support the workload-aware mapping in real-time. Experimental results demonstrate that \method~achieves 5.41$\times$ speedup over the edge GPU baseline on average and up to 17.3$\times$ speedup with the customized accelerator, while incurring only minimal visual quality degradation. 

\end{abstract}



\section{Introduction}
\label{sec:introduction}

Volume rendering is a fundamental technique for reconstructing and synthesizing novel views of 3D scenes, playing a crucial role in applications such as augmented/virtual reality (AR/VR) \cite{katragadda2024nerf,jiang2024vr} and autonomous driving \cite{cao2024lightning,tonderski2024neurad}. With the emergence of embodied intelligence \cite{qi2024air,huang2025enerverse,chen2025splat} and world models \cite{zhao2024drivedreamer4d,ni2024recondreamer}, 3D representation has become an essential modality for bridging the physical and virtual worlds. This shift imposes stringent requirements on both the quality and efficiency. However, traditional methods such as Neural Radiance Fields (NeRF) \cite{mildenhall2021nerf,li2022rt} are inadequate to meet these increasing demands, necessitating more efficient and scalable solutions. 

3D Gaussian Splatting (3DGS) \cite{kerbl20233d,chen2024survey,wu2024recent} models scenes as collections of Gaussian ellipsoids with varying shapes, colors, and opacities, synthesizing novel views through projection and accumulation of Gaussians onto the image plane. 
With explicit representations and tile-based parallel rasterization, 3DGS outperforms NeRF-based methods \cite{li2023instant,feng2024cicero,li2024fusion} in both quality and rendering speed, gaining significant attention since its introduction. 
However, as scene scales expand in practical applications \cite{liu2024citygaussian,fan2024momentum}, 3DGS must process millions or more Gaussians, incurring substantial memory and computational overhead. 
On resource-constrained edge platforms, such as the Jetson AGX Orin, which are critical for many real-world tasks, 3DGS still struggles to achieve real-time rendering at 90 FPS \cite{lin2025metasapiens}.

\begin{figure}[!tb]
    \centering
    \includegraphics[width=0.95\linewidth]{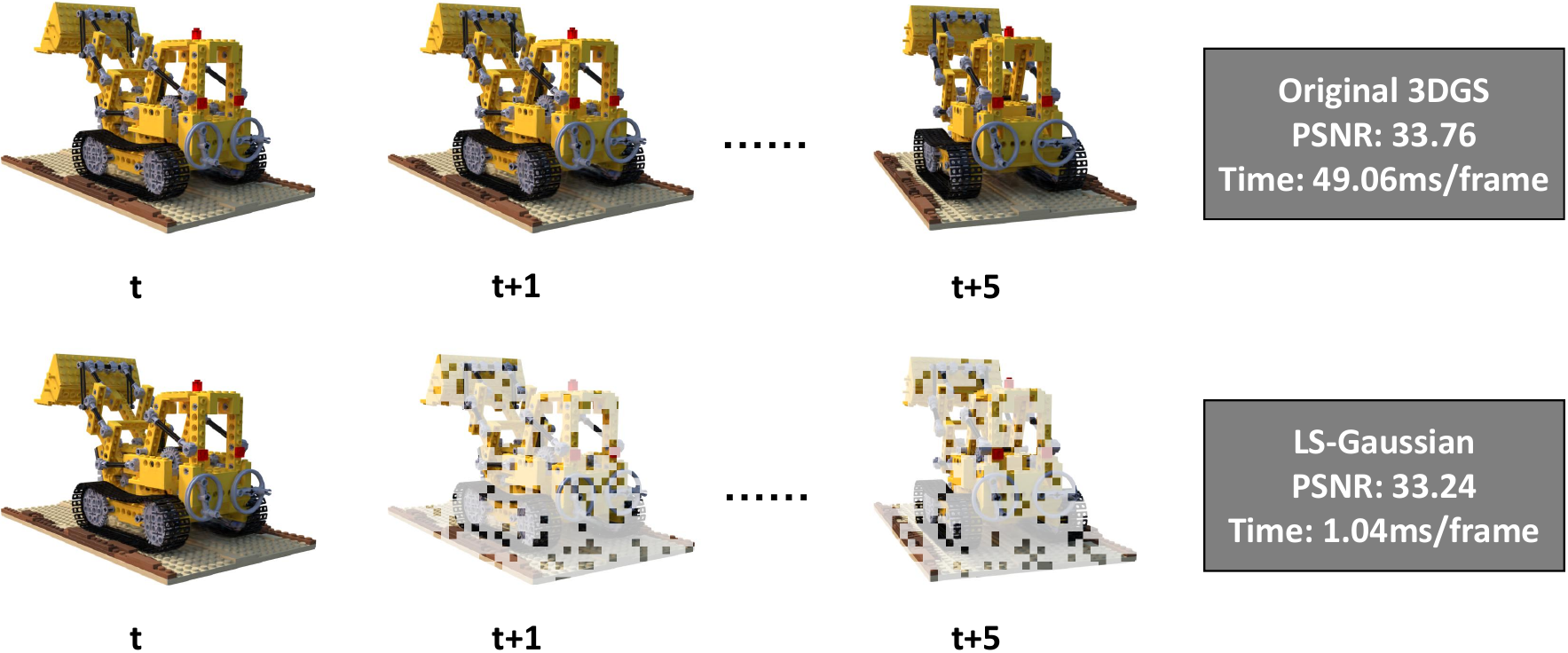}
    \caption{Rendering with 3DGS on Jetson AGX Orin and LS-Gaussian. LS-Gaussian is accelerated by sparse rendering between consecutive frames and only needs to fully render one in every 6 frames.}
    \label{Sparse rendering}
\end{figure}

Recent acceleration efforts focus on reducing Gaussian primitives and model size via quantization \cite{navaneet2024compgs,di2025gode}, pruning \cite{fang2024mini,ye20243d,zhang2024gaussianspa}, and level-of-detail (LOD) hierarchical rendering \cite{lu2024scaffold,ren2024octree}. 
However, these methods often degrade rendering quality and require costly retraining. 
Other approaches \cite{feng2024flashgs,huang2025seele,lee2024gscore,lin2025metasapiens} leverage hardware architecture innovations to enhance rendering efficiency but lack holistic pipeline optimization, limiting acceleration gains. 
More importantly, they overlook the high similarity between consecutive frames, missing opportunities to leverage prior frames for real-time rendering.


In this paper, we present a detailed analysis of bottlenecks across different stages of the 3DGS rendering pipeline, revealing inefficiencies caused by both computational redundancy and hardware stalls. The redundancy arises primarily from two sources. On one hand, 3DGS is often deployed in scenarios where an observer renders a scene from slightly shifting viewpoints across consecutive frames. This results in substantial inter-frame similarity, making it unnecessary to fully re-render every frame. On the other hand, coarse intra-frame intersection tests introduce a large number of invalid Gaussian-tile pairs into the rendering pipeline, further exacerbating redundant computation. In parallel, hardware stalls are mainly attributed to inter-block idling and intra-block bubbles, which stem from imbalanced and unpredictable tile workloads. The naive assignment of tiles to rendering blocks fails to achieve a balanced load distribution, leading to significant underutilization of hardware resources.

\begin{figure*}[!tb]
    \centering
    \includegraphics[width=0.95\linewidth]{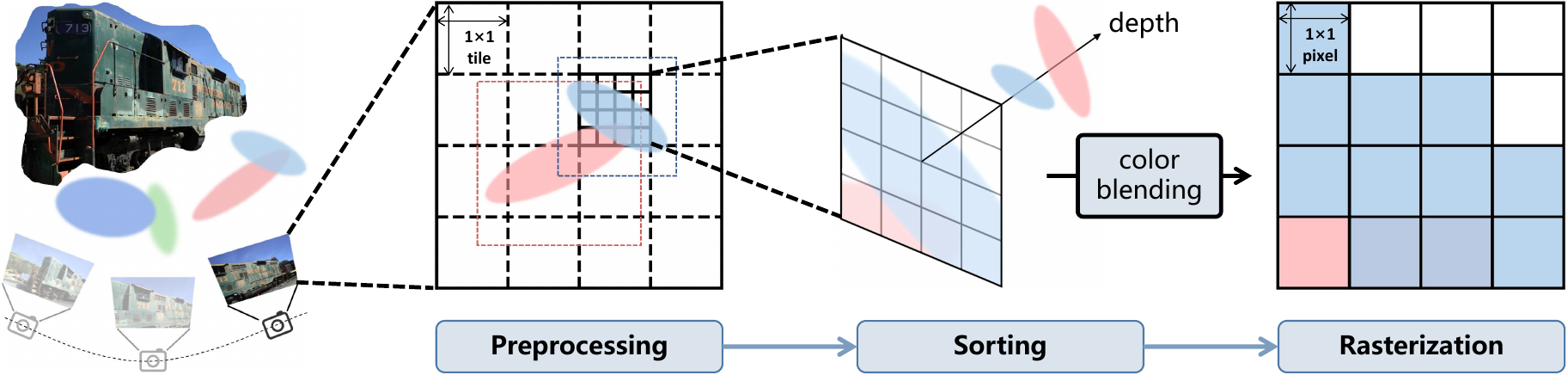}
    \caption{Overview of the 3DGS pipeline.}
    \label{3DGS pipeline}
\end{figure*}

Motivated by these observations, we propose \method, a 3DGS pipeline optimized for real-time rendering without retraining, leveraging frame continuity, as illustrated in \fref{Sparse rendering}. 
By refining the algorithm from inter-frame level to intra-frame level, we achieve redundancy-free, lightweight rendering. 
Additionally, we design a load-balanced accelerator architecture tailored for 3DGS, featuring a streaming pipeline that allows continuous processing across hardware modules without requiring global synchronization. This design improves hardware utilization and further enhances acceleration.
Our main contributions can be summarized as follows:

\begin{itemize}
    \item We conduct a comprehensive bottleneck analysis of the entire 3DGS pipeline, identifying two sources of redundancy and two types of stalls that limit rendering performance.
    \item To address these inefficiencies, we propose Lightweight Streaming 3DGS (\method), which optimizes the pipeline from both algorithmic and architectural perspectives. Our training-free approach enables seamless integration with existing methods.
    \item Extensive experiments across multiple datasets demonstrate that \method~achieves an average 5.41× speedup on Jetson AGX Orin GPU. With dedicated hardware support, \method~further achieves a 17.3× speedup while maintaining high quality.
\end{itemize}

\section{Background}
\label{sec:background}

\subsection{Preliminaries for 3D Gaussian Splatting}
3DGS represents a 3D scene using Gaussian ellipsoids, learned from a sparse point cloud generated by Structure from Motion (SfM) \cite{schonberger2016structure}. Each Gaussian is defined by its position, covariance matrix, opacity, and spherical harmonic coefficients, which encode its appearance color. During inference, the color of each pixel on the camera plane is computed by accumulating contributions from Gaussians projected onto that point. As illustrated in \fref{3DGS pipeline}, novel view synthesis in 3DGS consists of three stages:


\textbf{Preprocessing.} Gaussians outside the camera frustum are first culled and the remaining ones are projected onto the camera plane, which is subdivided into 16×16-pixel tiles. An axis-aligned bounding box (AABB) intersection test determines the tiles intersected by each Gaussian. It enables all pixels within the same tile to share common Gaussians, effectively leveraging the parallel computing capabilities of GPU streaming multiprocessors (SMs) during following steps.

\textbf{Sorting.} To accurately resolve occlusions from the viewpoint, Gaussians within each tile are sorted in depth order. This sorted sequence dictates the processing order during rasterization, ensuring correct color accumulation and blending.

\textbf{Rasterization.} SMs are organized into 16×16-thread blocks, where each tile is mapped to a unique block and each pixel within the tile is assigned to a distinct thread. During parallel rendering, threads within a block process the Gaussians covering the tile in Single Instruction Multiple Threads (SIMT) manner, following the sorted depth order. Each thread evaluates the density $\alpha_i$ of Gaussian $i$ with opacity $o_i$:
\begin{equation}\label{eq1}
\begin{aligned}
\alpha_i=o_i e^{-\frac{1}{2}\left(P-\mu_i'\right)^\top \Sigma_i'^{-1}\left(P-\mu_i'\right)},
\end{aligned}
\end{equation}
where $P$ represents the pixel coordinate, $\mu_i'$, $\Sigma_i'$ are the position and 2D covariance matrix of the projected Gaussian, respectively. If the density exceeds a predefined threshold (\(\frac{1}{255}\)), the Gaussian color $c_i$ accumulates, and the pixel color $C$ can be subsequently calculated through the volume rendering ($\alpha$-blending) process:
\begin{equation}\label{eq2}
\vspace{-4pt}
\begin{aligned}
C=\sum_{i \in N} c_i \alpha_i T_i=\sum_{i \in N} c_i \alpha_i \prod_{j=1}^{i-1}\left(1-\alpha_j\right).
\end{aligned}
\end{equation}
Once the cumulative transmittance $T_i$ drops below another threshold ($10^{-4}$), the pixel is considered fully rendered and thread execution is terminated, which is known as early stopping \cite{feng2024flashgs}.

\subsection{Acceleration of 3D Gaussian Splatting}
Several studies have highlighted the need to accelerate 3DGS. A common approach is to reduce the number of Gaussians and parameter size. Early efforts \cite{fang2024mini,ye20243d,zhang2024gaussianspa} explored quantization and pruning techniques, selectively removing Gaussians based on rendering sensitivity. Recently, the demand for reconstruction of city-scale scenes drives the adoption of level-of-detail (LOD) \cite{lu2024scaffold,ren2024octree,lin2025metasapiens}, which dynamically adjusts the Gaussian resolution based on depth and visual attention. However, these methods inevitably degrade reconstruction quality and rely on retraining. As 3DGS has become a core spatial representation, numerous studies have rapidly emerged to adapt it to new application scenarios. Retraining-based approaches require substantial modifications to accommodate different reconstruction paradigms, limiting their general applicability.

Some studies have explored accelerating 3DGS without retraining. Pixel and Gaussian reuse \cite{feng2024potamoi,tao2025gs} exploit the similarity between adjacent frames to alleviate computational and memory constraints, but fail to account for the tile-based rendering characteristics of 3DGS, resulting in limited speedup. Preprocessing techniques \cite{lee2024gscore,feng2024flashgs,wang2024adr,hanson2024speedy} improve the Gaussian-tile intersection accuracy, reducing unnecessary sorting and rendering, but often incur high computational costs due to complex analytical geometry. Meanwhile, GPU enhancements \cite{li2025gaurast,ye2025gaussian} and streaming designs \cite{feng2024flashgs,huang2025seele} aim to mitigate mutual stalls between hardware units, but lack fine-grained evaluation of workload imbalance. In contrast, we systematically address bottlenecks across the entire 3DGS pipeline, minimizing unnecessary computation while maximizing hardware utilization.

\section{Motivation}
\label{sec:motivation}

\begin{figure}[!tb]
    \centering
    \includegraphics[width=0.95\linewidth]{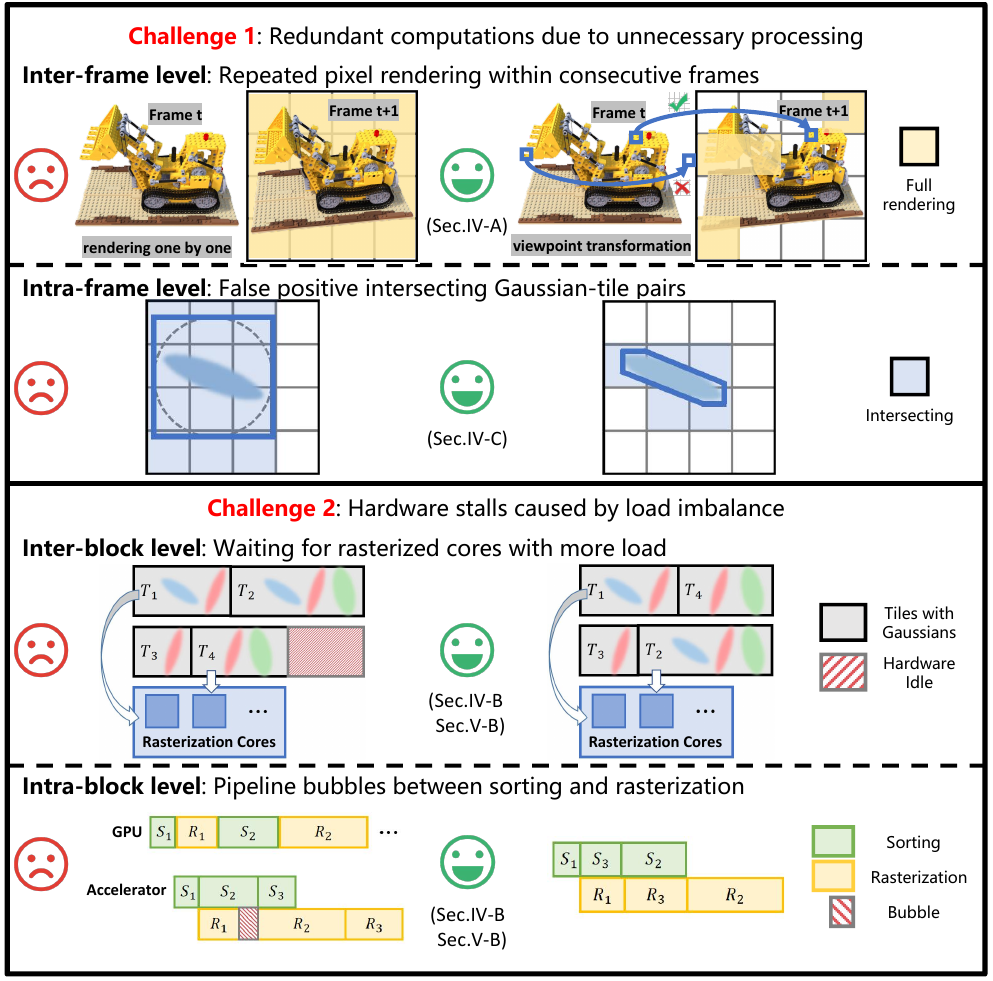}
    \caption{Algorithmic and hardware challenges of 3DGS.}
    \label{Challenges}
\end{figure}

To further investigate the performance bottlenecks in the 3DGS pipeline, we perform a comprehensive study of rendering speed and hardware efficiency across multiple datasets. Our findings indicate that inefficiencies primarily arise from redundant computations due to unnecessary processing and hardware stalls caused by load imbalance, as shown in \fref{Challenges}. 
Based on these insights, we present two key observations that motivate our subsequent work.

\begin{figure}[!tb]
    \centering    
    \subfloat[][]{
	\includegraphics[width=0.95\linewidth]{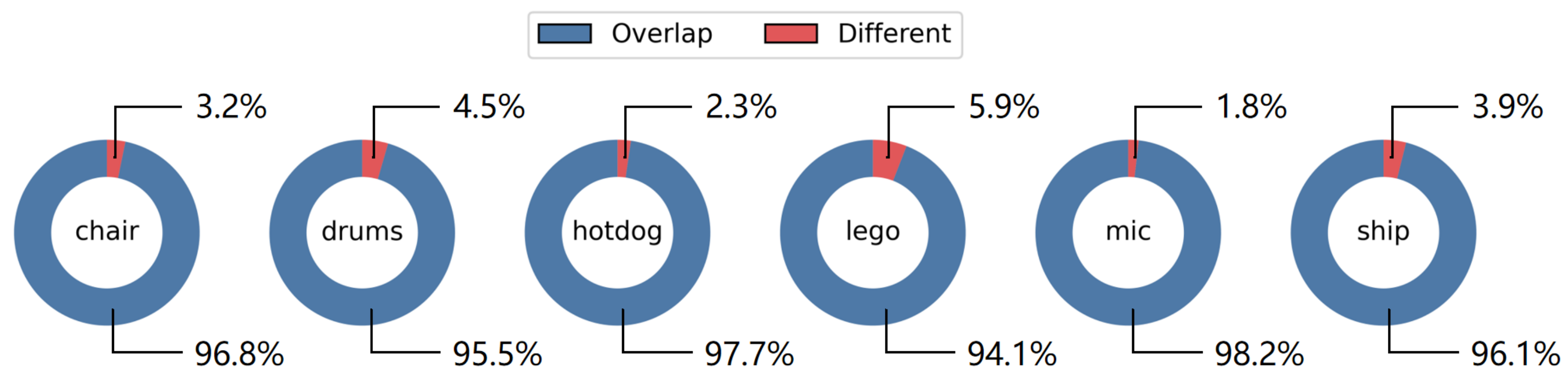}
        \label{Proportion of overlap pixels}
    }
    \hspace{0.01\textwidth}
    \subfloat[][]{
	\includegraphics[width=0.95\linewidth]{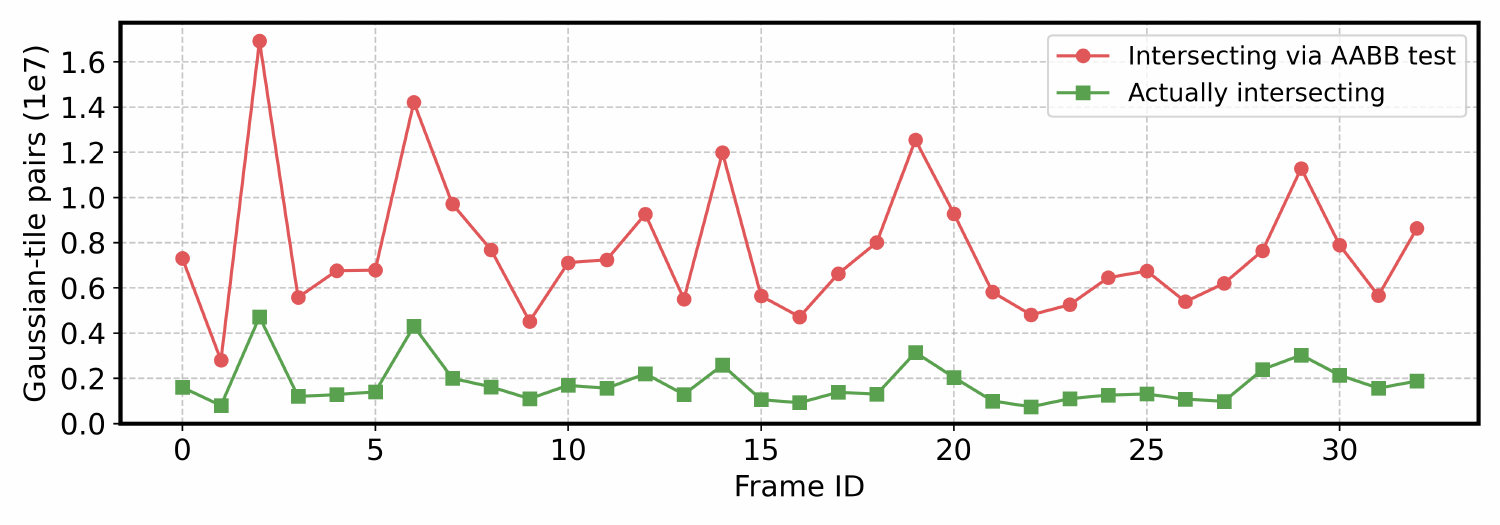}
        \label{Intersecting pairs}
    }
    \caption{(a) Proportion of overlap pixels between consecutive frames on multiple scenes and (b) the intersecting Gaussian-tile pairs judged by AABB test of the original 3DGS and actual intersecting pairs in the “drjohnson” scene test set.}
    \label{fig4}
\end{figure}

\textbf{Observation 1: Redundant computation exists across multiple levels of 3DGS rendering.} At the \textbf{inter-frame level}, real-time 3D rendering demands seamless transitions between frames, leading to substantial pixel overlap, especially in static or slowly changing regions, as illustrated in \fref{Proportion of overlap pixels}. 
We emphasize that existing efforts~\cite{feng2024potamoi} to handle viewpoint transformations in 3DGS remain inadequate. Since projected Gaussians are assigned to workloads at the tile level, naive pixel warping fails to reduce preprocessing and sorting costs. Moreover, rendering quality degrades noticeably as the number of consecutively transformed frames increases, due to the accumulation of interpolation errors. In addition, viewpoint transformations inherently provide depth priors that could be leveraged to predict early stopping positions, enabling more accurate tile load estimation for balanced task allocation. However, this potential remains largely unexploited.

At the \textbf{intra-frame level}, the original preprocessing stage overestimates the number of tiles covered by each Gaussian and increases invalid Gaussian-tile pairs to be rendered. However, as shown in \fref{Intersecting pairs}, this coarse bounding box approximation associates each tile with numerous Gaussians that do not actually contribute to rendering, significantly increasing the overhead in both the sorting and rasterization stages. Analytical geometry-based methods provides a more accurate intersection test but introduce significant overhead in the preprocessing stage.

\begin{figure}[!tb]
    \centering
    \includegraphics[width=0.95\linewidth]{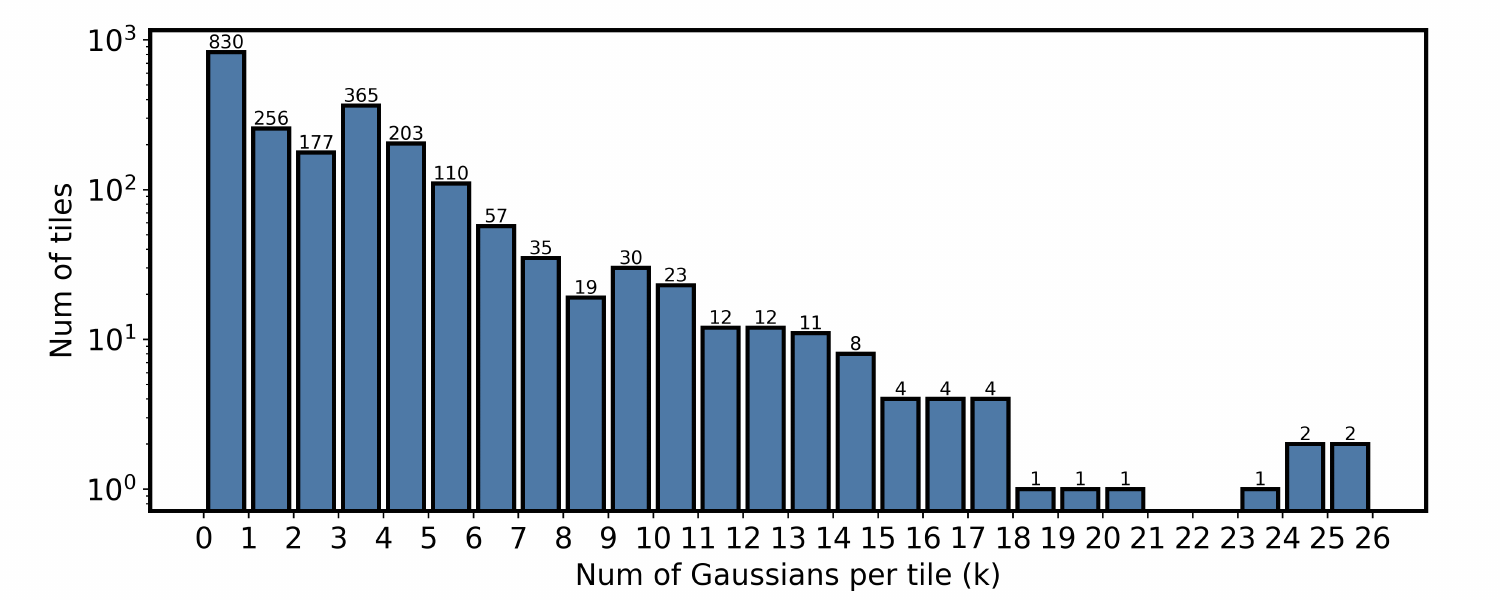}
    \caption{Distribution of the covered Gaussian numbers on different tiles in a frame from the “train” scene.}
    \label{Gaussian distribution}
\end{figure}

\textbf{Observation 2: Load imbalance causes inefficient hardware utilization.} As mentioned in \sref{sec:background}, tiles are mapped to streaming multiprocessor (SM) blocks for parallel rendering. However, as illustrated in \fref{Gaussian distribution}, where image tiles are grouped by the number of covered Gaussians, variations in scene complexity cause the per-tile Gaussian count to vary by more than an order of magnitude, leading to severe load imbalance.

At the \textbf{inter-block level}, this imbalance causes lightly loaded blocks to finish rasterization early but remain idle, waiting for heavily loaded blocks to complete rendering. On edge devices, additional processing waves due to limited compute resources help mitigate this imbalance by allowing underutilized blocks to render more tiles. However, sparse rendering which exploit frame-to-frame similarity reduce total workload and wave count, making inter-block waiting a critical bottleneck again. At the \textbf{intra-block level}, traditional GPU platforms perform sorting and rendering sequentially on SMs. Recent works such as GSCore \cite{lee2024gscore} enable parallelism between sorting and rasterization across tiles by decoupling these stages into dedicated hardware units, thereby reducing latency across them. However, the uncertainty in workload distribution results in potential bubbles in the rasterization unit, waiting for potentially prolonged sorting.



\section{Algorithm Optimization for Lightweight 3DGS}
\label{sec:algorithm}


Based on the insights from above analysis, we propose a series of algorithmic optimizations for lightweight 3DGS that eliminate computational redundancy at inter-frame and intra-frame levels. First, We introduce a novel sparse rendering strategy specifically designed for 3DGS (\sref{subsec:Sparse Rendering}), utilizing tile warping instead of the traditional pixel warping, thereby better aligning with the tile-based architecture of the pipeline. By leveraging depth map from viewpoint transformations, we then predict the locations where early stopping is likely to occur (\sref{subsec:Early Stopping}), effectively estimating the load of each tile. Additionally, we adopt a two-stage strategy to accurately determine the intersections between Gaussians and tiles (\sref{subsec:Intersection Test}).

\subsection{Tile Warpping-based Sparse Rendering (TWSR)}
\label{subsec:Sparse Rendering}
\begin{figure}[!tb]
    \centering
    \includegraphics[width=0.95\linewidth]{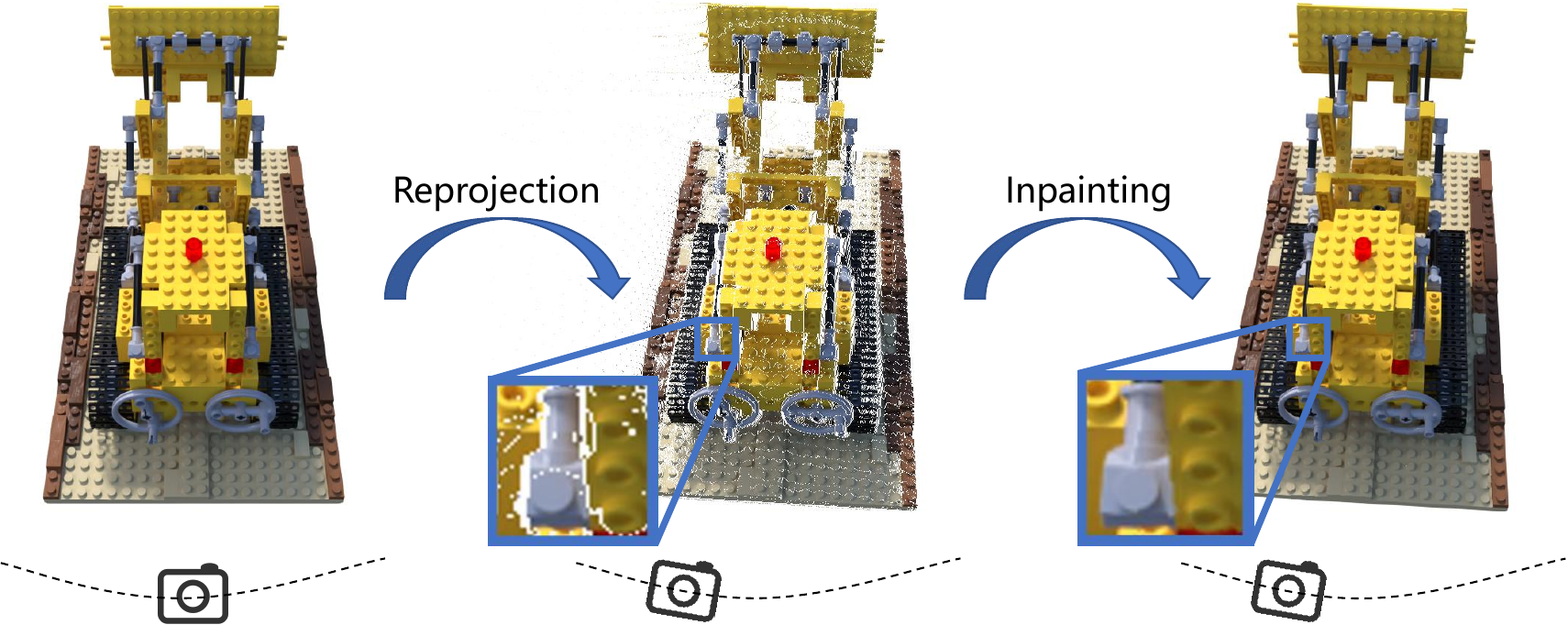}
    \caption{An example of viewpoint transformation and sparse rendering between consecutive frames. The reference frame is reprojected to the target viewpoint, and then inpaint the pixels that have no reprojection source.}
    \label{Viewpoint transformation}
\end{figure}

In real-time rendering scenarios, the high similarity between consecutive frames motivates the reuse of pixels to reduce redundant computations. As illustrated in \fref{Viewpoint transformation}, given a reference frame, we apply a viewpoint transformation to determine which pixels can be reused in the next frame. Specifically, pixels from the reference frame are first back-projected into 3D space using scene depth and camera pose, forming a point cloud. These 3D points are then transformed according to the target frame’s camera pose and reprojected onto its image plane. Due to viewpoint changes and resulting occlusions, the reprojected target frame typically contains missing pixels, which need to be filled through rendering.

\textbf{Pixel warping (PW).} Existing methods \cite{feng2024cicero,feng2024potamoi} often use a NeRF-style pixel warpping-based sparse rendering (PWSR) strategy, where only the missing pixels in the target frame are filled. However, we observe that this simple pixel warping approach conflicts with the 3DGS rendering pipeline. Unlike NeRF, where pixels are rendered independently, 3DGS organizes rendering by tiles, with all pixels in a tile sharing the same set of Gaussians. As a result, preprocessing and sorting cannot be skipped unless no pixel within a tile requires rendering. While rasterization remains the dominant factor in overall rendering time, sparse rendering reduces its workload, making the cost of preprocessing and sorting more pronounced. Moreover, unlike previous works that assume known ground truth depth, we estimate the depth in real-time by computing an opacity-weighted sum over the depths of the contributing Gaussians. While this estimation is generally accurate, it inevitably introduces minor errors. These small errors tend to be amplified in regions with limited pixel reuse, where view warping is less effective and missing pixels are more frequent. In such areas, the projected pixel positions deviate from the actual rendering results, ultimately degrading the final image quality.

\begin{figure}[!tb]
    \centering
    \includegraphics[width=0.95\linewidth]{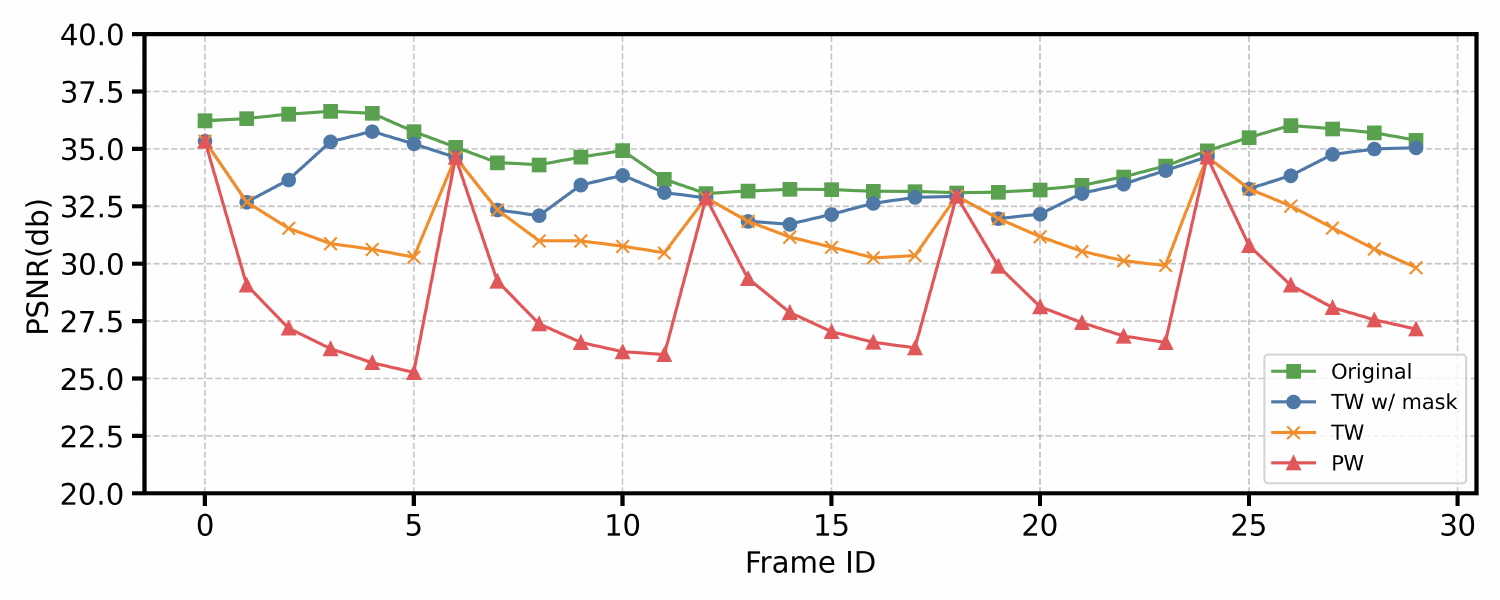}
    \caption{Comparison of image quality under different inpainting strategies for the “chair” scene in the Synthetic-NeRF dataset.}
    \label{Comparison of image quality}
\end{figure}

\textbf{Tile warping (TW).} To address the issues mentioned above, we propose a tile-based inpainting strategy. For tiles with a small number of missing pixels (empirically set to less than one-sixth of the total pixels in the tile), we observe that these typically correspond to regions with smooth depth variation and similar color distributions. In such cases, we directly interpolate the remaining pixels, bypassing not only the rasterization stage but also the preprocessing and sorting. On the other hand, for tiles with a large number of missing pixels, we re-render the entire tile to ensure high visual fidelity. As shown in \fref{Comparison of image quality}, our tile-level warping consistently achieves higher PSNR values compared to pixel-based warping approaches.

\textbf{No cumulative error mask (TW w/ mask).} Unfortunately, as the number of consecutive viewpoint transformations increases, we observe a noticeable degradation in image quality. This is primarily due to the accumulation of interpolation errors across frames. The more transformation rounds applied, the more severe the quality deterioration becomes. To mitigate this issue, we introduce a masking mechanism for interpolated pixels. These pixels are treated as "blank" during subsequent view projections and are excluded from contributing to the next frame. As shown in \fref{Comparison of image quality}, this masking strategy significantly enhances rendering quality compared to direct inpainting. Remarkably, the quality continues to improve with more consecutive frame transformations. This improvement is attributed to the increased number of missing pixels introduced by the mask in the target frame. According to our tile-based inpainting policy, this leads to a little more tiles being fully re-rendered, thereby striking a balance between efficiency and image quality.

\subsection{Depth Prediction for Early Stopping (DPES)}
\label{subsec:Early Stopping}
As mentioned in \sref{sec:motivation}, some tiles are covered by thousands of Gaussians, yet a significant portion of these Gaussians does not contribute to the color of the pixels in these tiles due to early stopping. 
As a result, the total number of overlapping Gaussians computed during preprocessing does not accurately reflect the actual workload per tile. To better estimate rendering cost, it is necessary to predict the early stopping position, which determines the number of Gaussians that are effectively traversed during rasterization.
In a typical rendering pipeline, this truncation point is unpredictable and depends on the opacity accumulation of Gaussians at each pixel during the rasterization stage. However, by leveraging viewpoint transformations based on scene depth and camera pose, we can predict and address this issue more effectively.

We emphasize that viewpoint transformations not only enable target frames to inherit pixel colors from reference frames but also their corresponding depths. Building on this, we propose a method to predict the early stopping point for each tile, leveraging the truncated depth from the reference frame. During the initial rendering, we record the depth of each pixel in the reference frame, which corresponds to the depth of the last Gaussian after traversing all Gaussians or the depth if early stopping occurs. These depths are reprojected onto the new camera plane. We define the early stopping depth for each tile in the target frame as the maximum depth of all valid reprojected pixels within that tile. Any Gaussians beyond this depth will not be involved in sorting in the target frame. The complete viewpoint transformation process, including sparse rendering and truncated depth estimation, is outlined in \aref{alg1}.

\begin{algorithm}[!tb]
\caption{Viewpoint Transformation Process}
\renewcommand{\algorithmicrequire}{\textbf{Input:}}
\renewcommand{\algorithmicensure}{\textbf{Output:}}
\label{alg1}
\begin{algorithmic}[1]
\REQUIRE A reference frame $F_{\text{ref}}$, scene depth map $D_{\text{ref}}$, truncated depth map $D_{\text{ref}}^{\text{max}}$, reference viewpoint $V_{\text{ref}}$, target viewpoint $V_{\text{tgt}}$, re-rendering threshold $N_0$
\ENSURE A target frame $F_{\text{tgt}}$ with a groups of tiles $T$, tile-level truncated depth map $D_{\text{T}}$ 
\STATE $F_{\text{tgt}}, D_{\text{T}} \gets \text{Initialize}()$
\STATE $P_{\text{ref}}, P_{\text{ref}}^{\text{max}} \gets \text{ProjectTo3D}(F_{\text{ref}}, D_{\text{ref}}, D_{\text{ref}}^{\text{max}})$
\STATE $P_{\text{tgt}}, P_{\text{tgt}}^{\text{max}} \gets \text{ViewTransfer}(P_{\text{ref}}, P_{\text{ref}}^{\text{max}})$
\STATE $F_{\text{tgt}}, D_{\text{tgt}}^{\text{max}} \gets \text{Reproject}(P_{\text{tgt}}, P_{\text{tgt}}^{\text{max}})$
\FOR{$t \in T$}
    \STATE $N \gets \text{ValidPixelSum}(F_{\text{tgt}}, t)$
    \IF{$N > N_0$}
        \STATE \text{Interpolate($F_{\text{tgt}}, t$)}
    \ELSE
        \STATE $D_{\text{T}} \gets \text{Max}(D_{\text{tgt}}^{\text{max}}, t)$ 
        \STATE \text{Re-render($F_{\text{tgt}}, t$)}
    \ENDIF
\ENDFOR
\end{algorithmic}
\end{algorithm}


The early stopping estimation can be utilized to predict the expected load of each tile after sorting, allowing for a more balanced distribution of the workload across rasterization units. This significantly mitigates kernel idle caused by load imbalance, as discussed in the next section.

\subsection{Two-stage Accurate Intersection Test (TAIT)}
\label{subsec:Intersection Test}
\begin{figure}[!tb]
    \centering
    \includegraphics[width=0.95\linewidth]{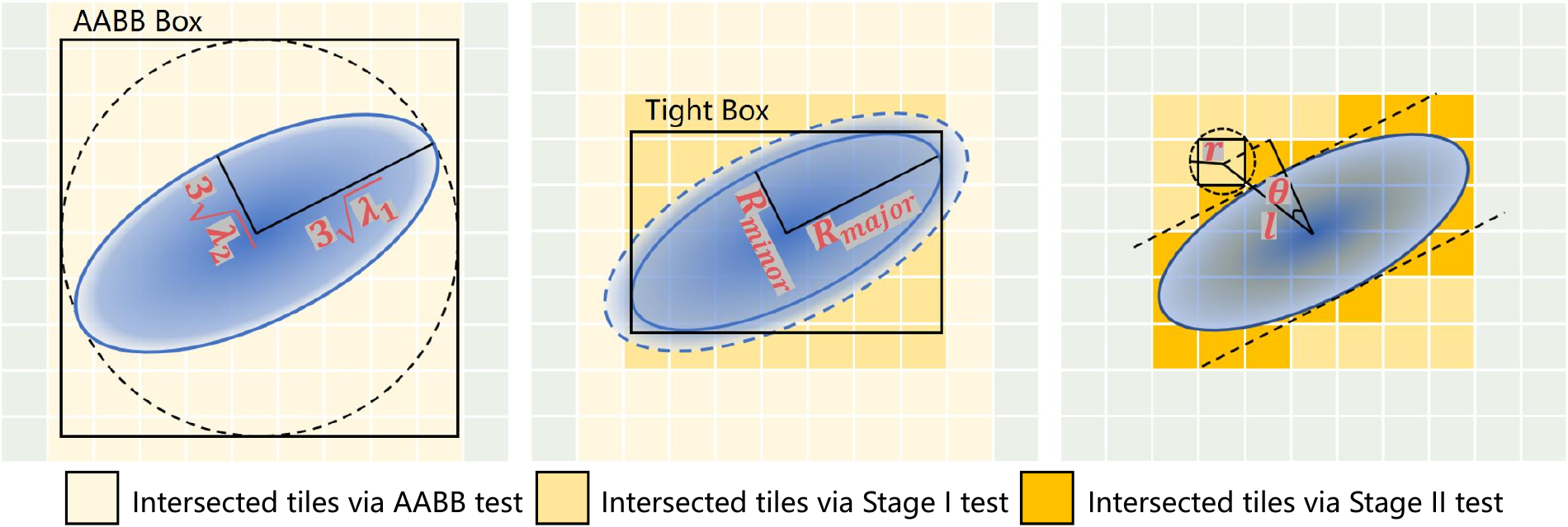}
    \caption{Two-stage accurate intersection test.}
    \label{Intersection test}
\end{figure}

As mentioned in \sref{sec:background}, a simple AABB test is commonly used to determine which tiles are covered by each Gaussian. This approach is efficient for preprocessing, as it only requires the projected center and the semi-major axis length of the Gaussian. The bounding box is constructed as the circumscribed square of a circle centered at the Gaussian's projection, with a radius equal to the semi-major axis. All tiles intersecting with this square are considered to intersect with the Gaussian. However, as shown in \fref{Intersection test}, this approximation leads to a significant number of false positives, greatly increasing the burden in the subsequent sorting and rasterization stages. This overestimation is especially pronounced for elongated Gaussians.

\begin{figure}[!tb]
    \centering
    \includegraphics[width=0.95\linewidth]{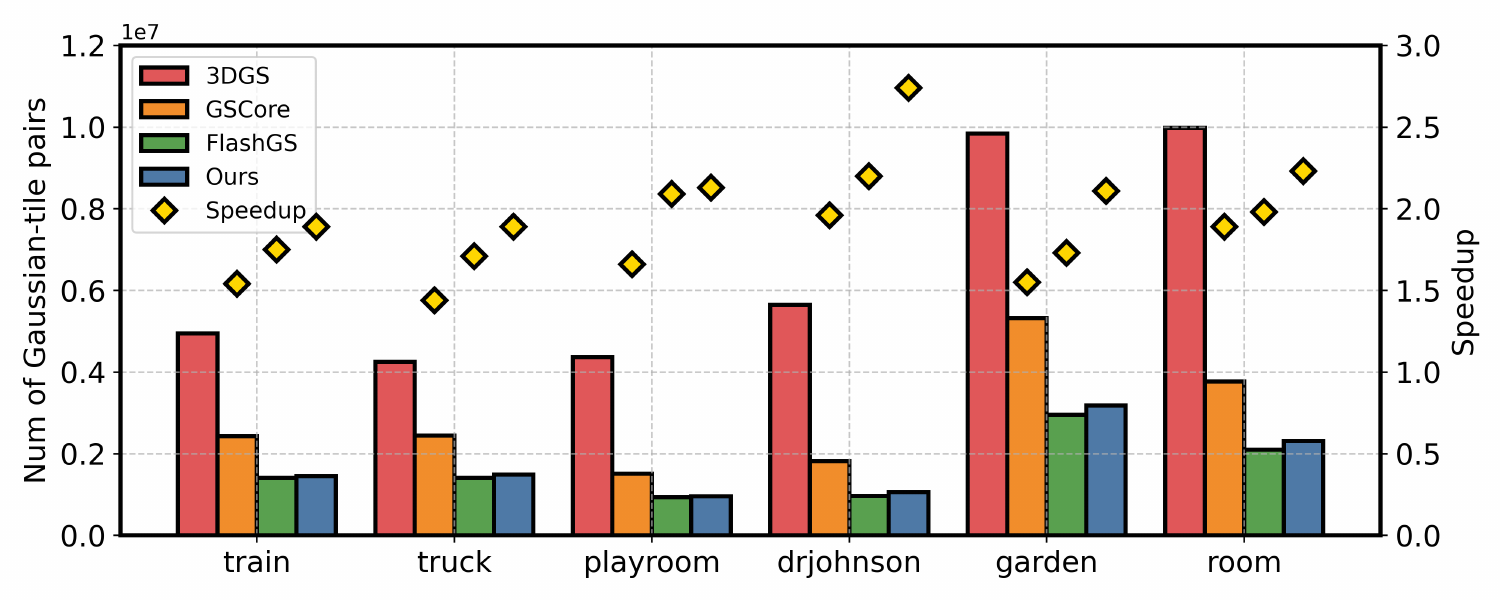}
    \caption{Comparison of our two-stage accurate intersection test with previous works on the number of Gaussian-tile pairs and Speedup across multiple scenes.}
    \label{Intersection speedup}
\end{figure}

We identify three primary sources of false-positive intersections: 1) The semi-major and semi-minor axes of each projected Gaussian are heuristically defined as \(3\sqrt{\lambda_1}\) and \(3\sqrt{\lambda_2}\), where $\lambda_1$ and $\lambda_2$ ($\lambda_1>\lambda_2$) are the eigenvalues of the 2D covariance matrix. However, in the rasterization stage, only Gaussians with opacity above the threshold contribute to rendering, reducing the effective coverage; 2) The standard approximation uses a circle with radius equal to the semi-major axis to bound the projected Gaussian, which significantly overestimates coverage along the minor axis. The use of a circumscribed square around this circle further inflates the bounding area, leading to excessive tile assignments; 3) Even within the tight bounding box of the ellipse, the geometric mismatch between the elliptical shape of the Gaussian and the rectangular shape of the box results in additional false positives.

To determine the intersection between Gaussians and tiles accurately, we propose a coarse-to-fine two-stage intersection test. In the first stage, our objective is to approximate a tight bounding box for each Gaussian at minimal computational cost, thereby filtering out the majority of non-overlapping Gaussian-tile pairs. We begin by modeling the opacity falloff of each Gaussian as a function of distance $d$ from its projected center:
\begin{equation}\label{eq3}
\begin{aligned}
\alpha_i=o_i e^{-{d^2} / {2 \lambda}}.
\end{aligned}
\end{equation}
Based on this formulation, we define the effective semi-major and semi-minor axes lengths of the projected Gaussian as the distances at which its opacity decays to the threshold $\tau=\frac{1}{255}$:
\begin{equation}\label{eq4}
\begin{aligned}
R_{major}=\sqrt{2 \ln \left(\frac{o_i}{\tau}\right) \lambda_1} ,\quad R_{minor}=\sqrt{2 \ln \left(\frac{o_i}{\tau}\right) \lambda_2}.
\end{aligned}
\end{equation}

Similar to the analysis in \cite{wang2024adr,hanson2024speedy}, the boundaries of a tight bounding box for an ellipse can be determined by locating the extrema points of the ellipse equation $F(x,y)$ along the Cartesian axes, where the partial derivatives with respect to x and y are zero:
\begin{equation}\label{eq5}
\begin{aligned}
\frac{\partial F(x,y)}{\partial x}=0 ,\quad \frac{\partial F(x,y)}{\partial y}=0.
\end{aligned}
\end{equation}
By substituting these extrema points into the elliptic equation, the width $W$ and height $H$ of the tight bounding box are obtained through semi-major axis length $R_{major}$ and components of the projection covariance matrix $\Sigma_X^{\prime}$, $\Sigma_Y^{\prime}$ in the x and y directions:
\begin{equation}\label{eq6}
\begin{aligned}
W=2\sqrt{\frac{\Sigma_X^{\prime}}{\lambda_1}}R_{major} ,\quad H=2\sqrt{\frac{\Sigma_Y^{\prime}}{\lambda_2}}R_{major}.
\end{aligned}
\end{equation}

\begin{figure*}[!tb]
    \centering
    \includegraphics[width=0.95\linewidth]{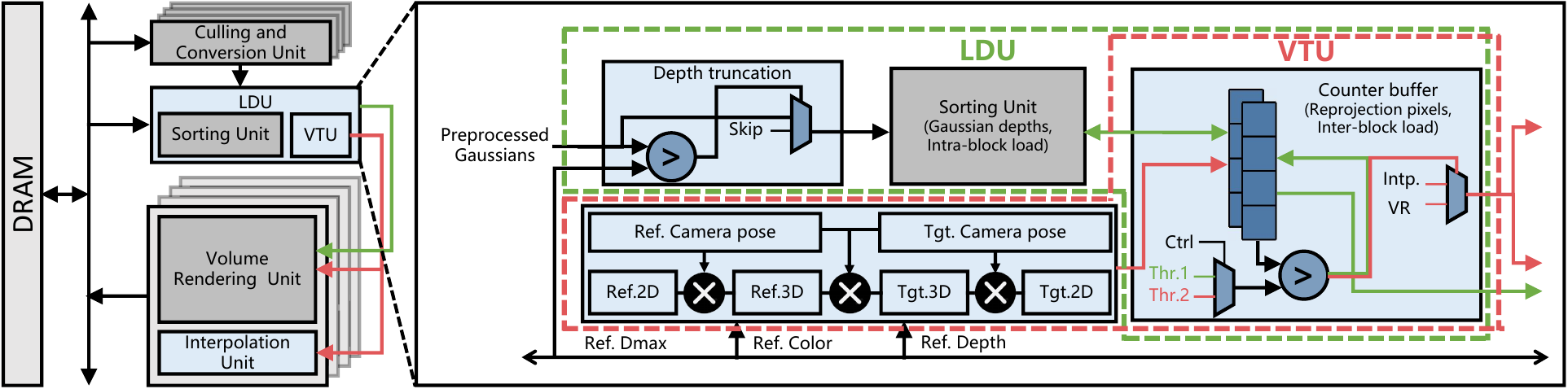}
    \caption{Overall design of LS-Gaussian architecture. Components inherited from GSCore \cite{lee2024gscore} are shown in gray, while newly introduced units are highlighted in blue. The augmented Viewpoint Transformation Unit (VTU) and Load Distribution Unit (LDU) are enclosed in red and green boxes, respectively, with colored data streams demonstrating the efficient hardware reuse capability of LS-Gaussian.}
    \label{LS-Gaussian architecture}
\end{figure*}

In the second stage, we further discard the tiles that do not actually intersect with the Gaussian within the tight bounding box. To strike a balance between accuracy and efficiency, and to avoid making the preprocessing stage a new bottleneck with complex analytical geometry, we consider $l$, the line connecting the center of each tile to the center of the ellipse onto the short axis of the ellipse, as shown in \fref{Intersection test}. We classify a tile as non-intersecting with the ellipse when:
\begin{equation}\label{eq7}
\begin{aligned}
|l| \cos \theta +r>R_{minor},
\end{aligned}
\end{equation}
where $\theta$ represents the angle between $l$ and semi-minor axis of the ellipse, $r$ represents the circumcircle radius of tiles. By simply computing and comparing the distance once, we are able to eliminate almost all false intersection pairs with minimal computational cost. As illustrated in \fref{Intersection speedup}, compared to prior intersection tests, our method retains substantially fewer Gaussian-tile pairs than rough screening approaches, such as GSCore's OOBB bounding box \cite{lee2024gscore}. At the same time, it introduces only a negligible amount of redundancy when compared to fully accurate intersection tests like FlashGS \cite{feng2024flashgs}, while significantly reducing the computational cost of complex geometric operations. As a result, our approach achieves optimal speedup across a variety of scenes.


\section{Streaming Hardware Architecture}
\label{sec:architecture}

Though it is undeniable that existing GPU platforms provide strong support for the parallel rendering, we identify the partial mismatch between GPU architectures and the 3DGS pipeline. Every stage of rendering heavily relies on GPU SMs, leading to frequent global memory accesses for reading and writing intermediate data. Also, different stages exhibit distinct computational characteristics and hardware demands, constraining rendering efficiency and hardware utilization. To further accelerate 3DGS, we design a dedicated hardware accelerator, \method, beyond the algorithmic optimizations above. The streaming mechanism enables early stages to initiate processing for subsequent frames while later stages are still executing previous ones, thereby improving parallelism and pipeline efficiency. We present the overall design (\sref{subsec:Overall Design}) and incorporate architectural implementation at both inter-block and intra-block (\sref{subsec:Load distribution unit}) levels to minimize stalls across processing units.

\subsection{Overall Design}
\label{subsec:Overall Design}
\fref{LS-Gaussian architecture} illustrates the overall architecture of \method~built upon GSCore \cite{lee2024gscore}, which comprises dedicated hardware units for preprocessing, sorting, and rasterization to enable better inter-tile overlap. On top of the original design, we enhance the Culling and Conversion Unit (CCU) responsible for Gaussian preprocessing to support our proposed two-stage accurate intersection test. This enhancement introduces a square root and logarithmic operator, while eliminating the need for GSCore’s complex dual OBB Intersection Test Units (OIUs).


In addition, we enhance the Viewpoint Transformation Unit (VTU) to support sparse rendering, and extend the Load Distribution Unit (LDU) to handle task assignment and execution scheduling for each rasterization block. These enhanced components are marked in color in \fref{LS-Gaussian architecture}. For each viewpoint, pixel colors and maximum depths from the reference frame are retrieved for view transformation. This process involves three matrix multiplications: mapping 2D pixels to 3D space, transforming the resulting point cloud, and reprojecting it onto the new camera plane. Given that the number of pixels is substantially smaller than the number of Gaussians, this transformation incurs negligible overhead and can be parallelized with preprocessing to fully hide its latency. A counter array tracks the number of validly projected pixels per tile. When this number exceeds $\frac{5}{6}$ of the total pixels within the tile, the tile is interpolated as described in \sref{sec:algorithm}. Otherwise, it is forwarded to the Volume Rendering Unit (VRU) for full re-rendering. We will discuss how we balance the workload among rasterization units for rendering these tiles.

\subsection{Load Distribution Unit}
\label{subsec:Load distribution unit}
Due to the severe imbalance in tile workloads, overall hardware utilization remains suboptimal. To address this, we leverage early stopping depth prediction to implement balanced workload distribution across rasterization blocks and adopt a light-to-heavy execution order within each block. After preprocessing, Gaussians that lie beyond its predicted early stopping depth are discarded, and the remaining ones are treated as the tile’s effective workload. We first compute the ideal average workload $W$ and assign approximately $N$ tiles per block. Tiles are then assigned to blocks sequentially. If the cumulative number of Gaussian-tile pairs in the current block exceeds \((1+\frac{1}{N})W\), the current tile is deferred to the next block. To further enhance memory efficiency, we employ a Morton-order (Z-order) traversal strategy, which groups spatially adjacent tiles into the same block to reduce memory access overhead. Notably, this inter-block workload distribution is executed in parallel with the sorting process to avoid introducing additional latency. Considering that the Viewpoint Transformation Unit (VTU) operates in parallel with the preprocessing stage of the original pipeline, the counter array and comparators within the VTU can be directly reused, introducing no additional hardware overhead.

After completing the workload distribution, tiles within each block are sorted based on their workload, from light to heavy. Since the sorting process typically takes less time than rasterization, this ordering ensures that each tile can quickly access the pre-sorted Gaussians during the rasterization stage. It provides sufficient time for sorting high-workload tiles, preventing long sorting times from exceeding the rasterization time of lighter workload tiles, thus avoiding rasterization bubbles and reducing overall rendering time. The overhead of sorting the tiles’ workloads is minimal in comparison to the Gaussian depth sorting. Therefore, we can reuse the Gaussian Sorting Unit (GSU) for this task, similar to the inter-block workload distribution design, without the need for additional hardware units.
\section{Experimental Results}
\label{sec:experiments}

\subsection{Experimental Setup}
\textbf{Benchmarks.} We evaluate our rendering pipeline on the Synthetic-NeRF dataset \cite{mildenhall2021nerf}, which includes eight scenes with continuous viewpoints. To further validate the generality of our approach, we also select six real world scenes, including three indoor scenes (playroom, drjohnson, room) and three outdoor scenes (train, truck, garden), from three widely adopted datasets: Tanks \& Temples \cite{knapitsch2017tanks}, Deep Blending \cite{hedman2018deep}, and Mip-NeRF 360 \cite{barron2022mip}. Since these real world datasets typically contain sparse camera trajectories, we perform interpolation to construct continuous view sequences suitable for real time rendering at 90 FPS. This setup simulates camera motion at 1.8 meters and a rotational speed of 90 degrees per second.

\textbf{Baselines.} We use the original 3DGS running on Jetson AGX Orin as our GPU baseline. To comprehensively evaluate the effectiveness of our proposed \method~in both algorithmic optimization and architectural implementation, we compare it with five recent 3DGS acceleration methods. Specifically, we evaluate rendering quality against Potamoi \cite{feng2024potamoi}, another method based on viewpoint transformation. For performance evaluation, we compare GPU acceleration with AdR-Gaussian \cite{wang2024adr} and SeeLe \cite{huang2025seele}, and compare our accelerator design with GSCore \cite{lee2024gscore} and MetaSapiens \cite{lin2025metasapiens}.

\textbf{Hardware Implementation.} We implement the RTL design of \method~hardware components using Synopsys synthesis in 16nm FinFET technology. Based on GSCore, we replace part of the multipliers and adders in the Culling and Conversion Unit (CCU) with a square root and logarithm operator. In addition, we introduce an interpolation unit, a 16KB counter buffer and other specially designed units. Benefiting from the reuse of existing hardware modules by the Load Distribution Unit (LDU), our total design area is only 1.84 mm\textsuperscript{2}, representing just a 0.39 mm\textsuperscript{2} increase over the scaled GSCore design in 16nm (1.45mm\textsuperscript{2}), and remaining significantly smaller than Jetson-series edge GPUs (\raisebox{-0.5ex}{\textasciitilde}350 mm\textsuperscript{2}) and MetaSapiens (2.73 mm\textsuperscript{2}).

\subsection{Rendering Quality}
\begin{figure}[!tb]
    \centering    
    \subfloat[][]{
	\includegraphics[width=0.95\linewidth]{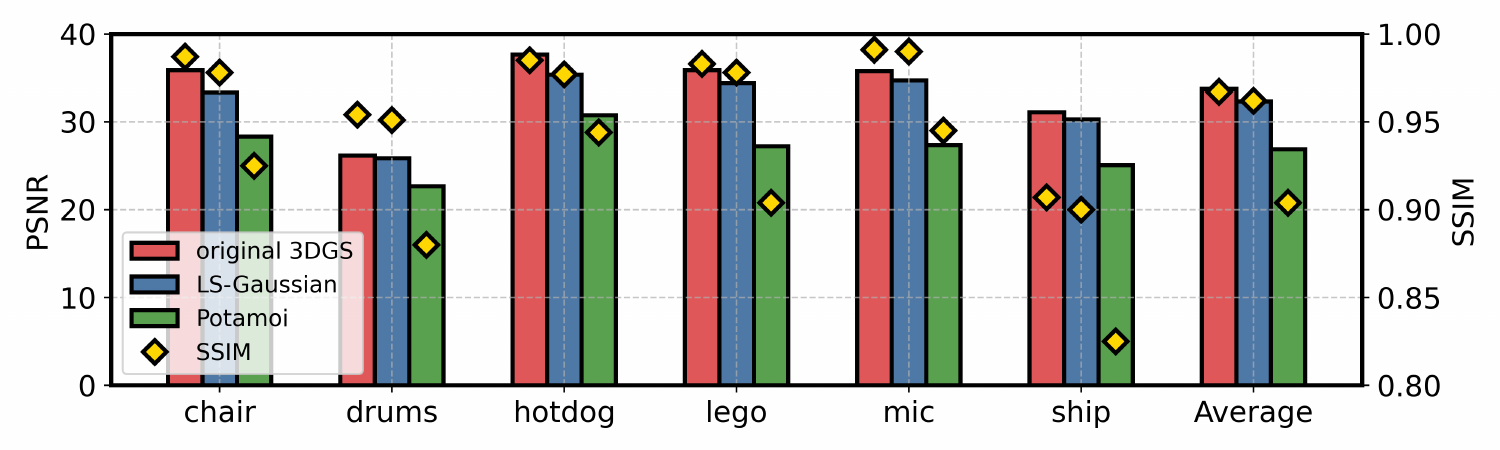}
        \label{Synthetic rendering quality}
    }
    \hspace{0.01\textwidth}
    \subfloat[][]{
	\includegraphics[width=0.95\linewidth]{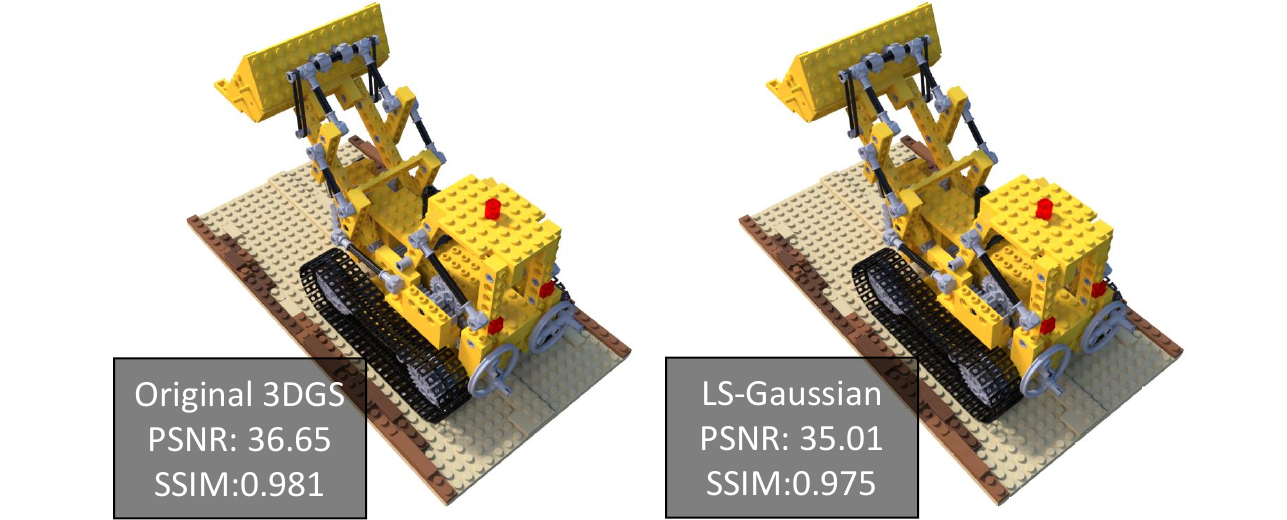}
        \label{Synthetic visual quality}
    }
    \caption{(a) Comparison of the rendering quality between LS-Gaussian and original 3DGS \cite{kerbl20233d}/ Potamoi \cite{feng2024potamoi} on the Synthetic-NeRF dataset and (b) two representative rendering output images with original 3DGS and our LS-Gaussian, respectively.}
    \label{fig11}
\end{figure}

We compare the rendering quality of our Tile Warping-based Sparse Rendering (TWSR) against original 3DGS and Potamoi on the Synthetic-NeRF dataset, as shown in \fref{Synthetic rendering quality}. Both TWSR and Potamoi adopt a strategy of fully rendering one frame every six frames. Our method results in an average SSIM loss of only 0.005 and a PSNR loss of 1.4 dB across six scenes, which is significantly lower than Potamoi's SSIM loss of 0.063 and PSNR loss of 6.8 dB. This is because Potamoi’s pixel-based inpainting ignores potentially invalid reprojections when depth priors are unavailable, resulting in incorrect floating pixels and severe visual artifacts, whereas TWSR performs tile-level full re-rendering for key regions with insufficient reprojection. In addition, due to the need for full preprocessing and sorting, Potamoi achieves limited speedups. \fref{Synthetic visual quality} presents the visual result of a representative frame. Given that PSNR values on the Synthetic-NeRF dataset are typically above 30 dB, even more than 1.5 dB drop in TWSR does not introduce visible distortion.

\begin{figure}[!tb]
    \centering    
    \subfloat[][]{
	\includegraphics[width=0.95\linewidth]{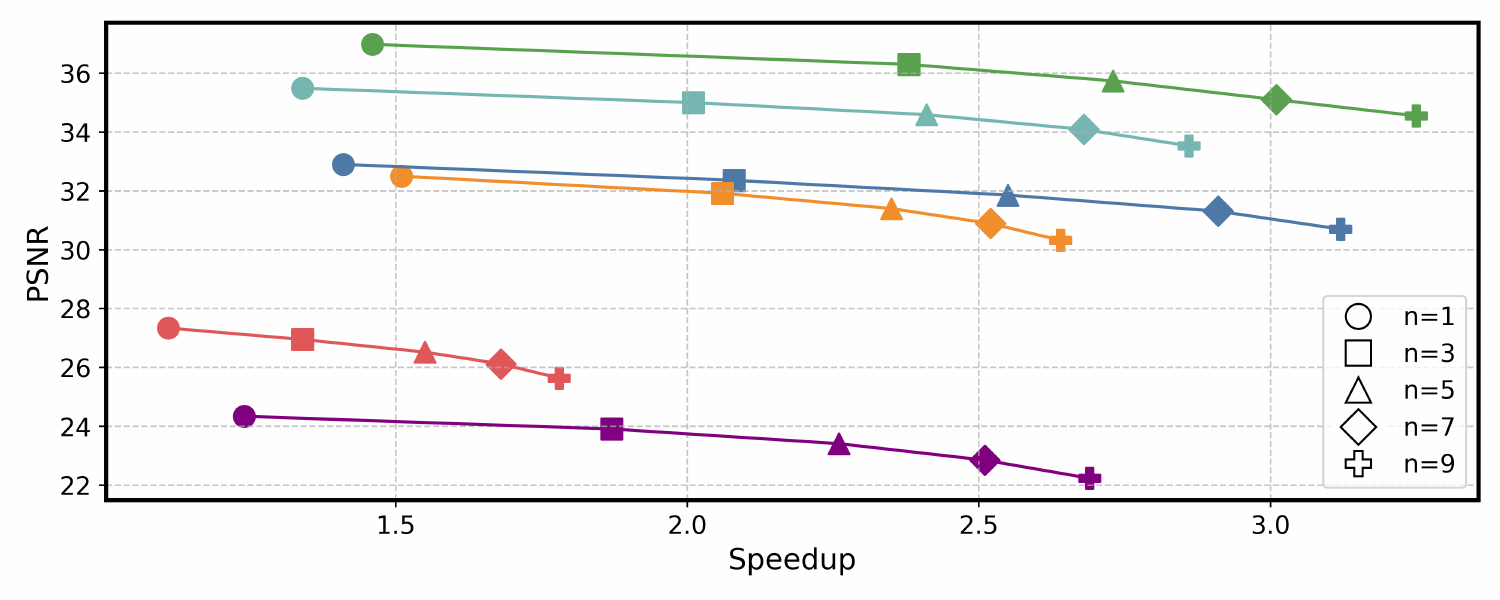}
        \label{Real speedup and PSNR}
    }
    \hspace{0.01\textwidth}
    \subfloat[][]{
	\includegraphics[width=0.95\linewidth]{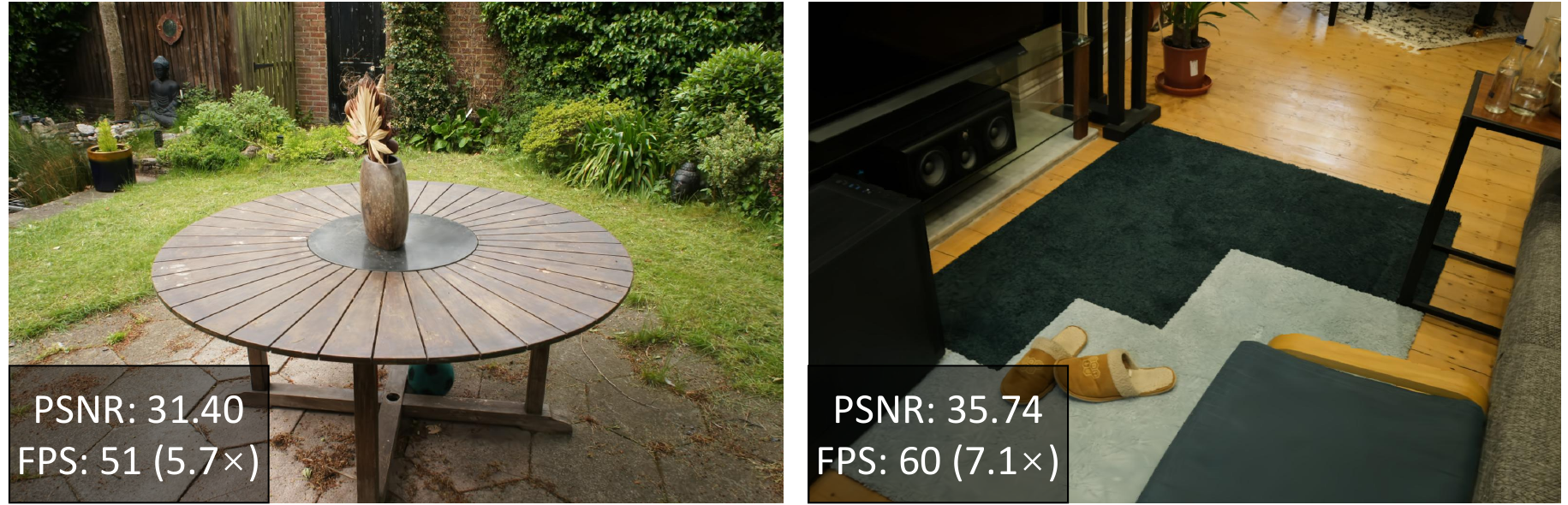}
        \label{Real visual quality}
    }
    \caption{(a) Speedup and PSNR (w/ and w/o TWSR) under different warping window size $n$ on multiple real-world scenes, where each color represents a scene and (b) two representative real-world rendering images w/ TWSR, tested on Jetson AGX Orin.}
    \label{fig12}
\end{figure}

We further investigate the sensitivity of image quality and rendering speedup to the warping window size $n$, which defines how many frames between every two fully rendered frames, on real-world scenes. As discussed earlier, we interpolate frames in the original real-world datasets to ensure frame continuity. Therefore, we evaluate image quality by measuring the difference between rendered outputs w/ and w/o the proposed viewpoint transformation. As shown in \fref{Real speedup and PSNR}, increasing the window size leads to greater speedups but also results in image quality degradation. To balance rendering quality and efficiency, we set $n=5$ as the default configuration for subsequent performance evaluations. The visual results and speedups of two representative frames are as shown in  \fref{Real visual quality}. We observe that indoor scenes, which typically exhibit smaller depth variations, achieve both better rendering quality and higher speedups compared to outdoor scenes with more edges and corners. This phenomenon will be discussed in detail in the following subsection.


\subsection{Performance on GPU}
\begin{figure}[!tb]
    \centering    
    \subfloat[][]{
	\includegraphics[width=0.95\linewidth]{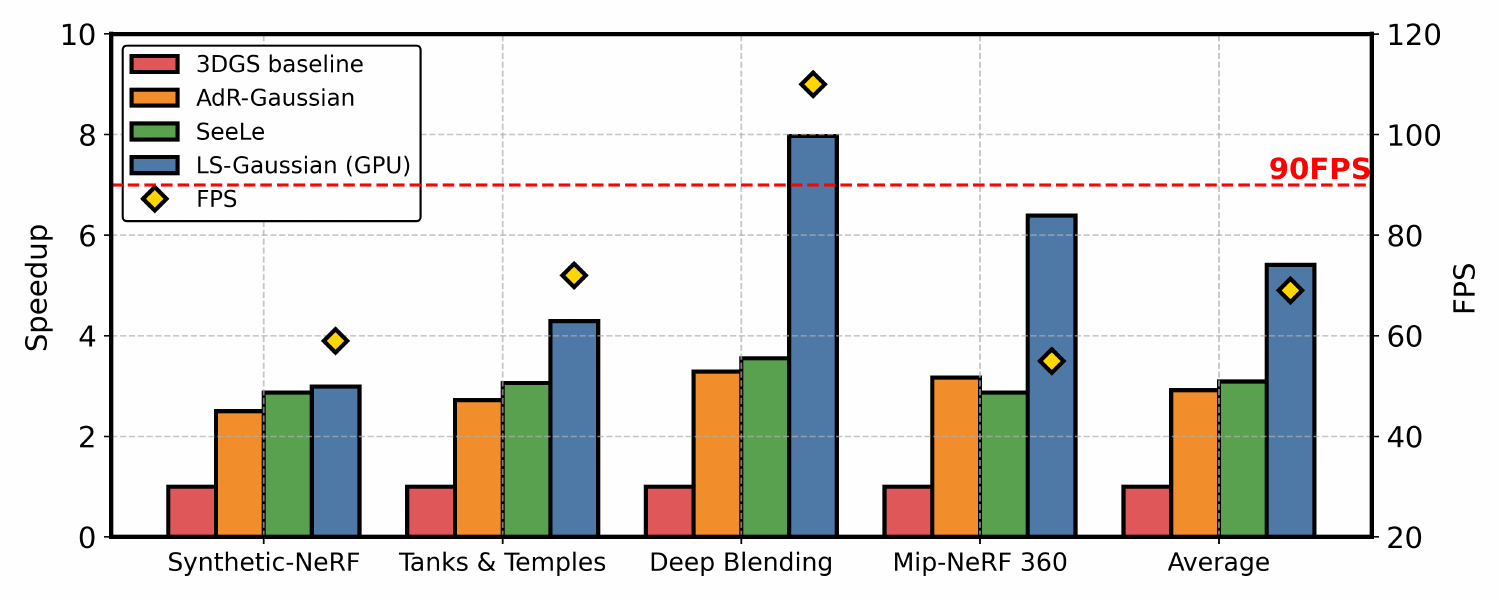}
        \label{GPU performance}
    }
    \hspace{0.01\textwidth}
    \subfloat[][]{
	\includegraphics[width=0.95\linewidth]{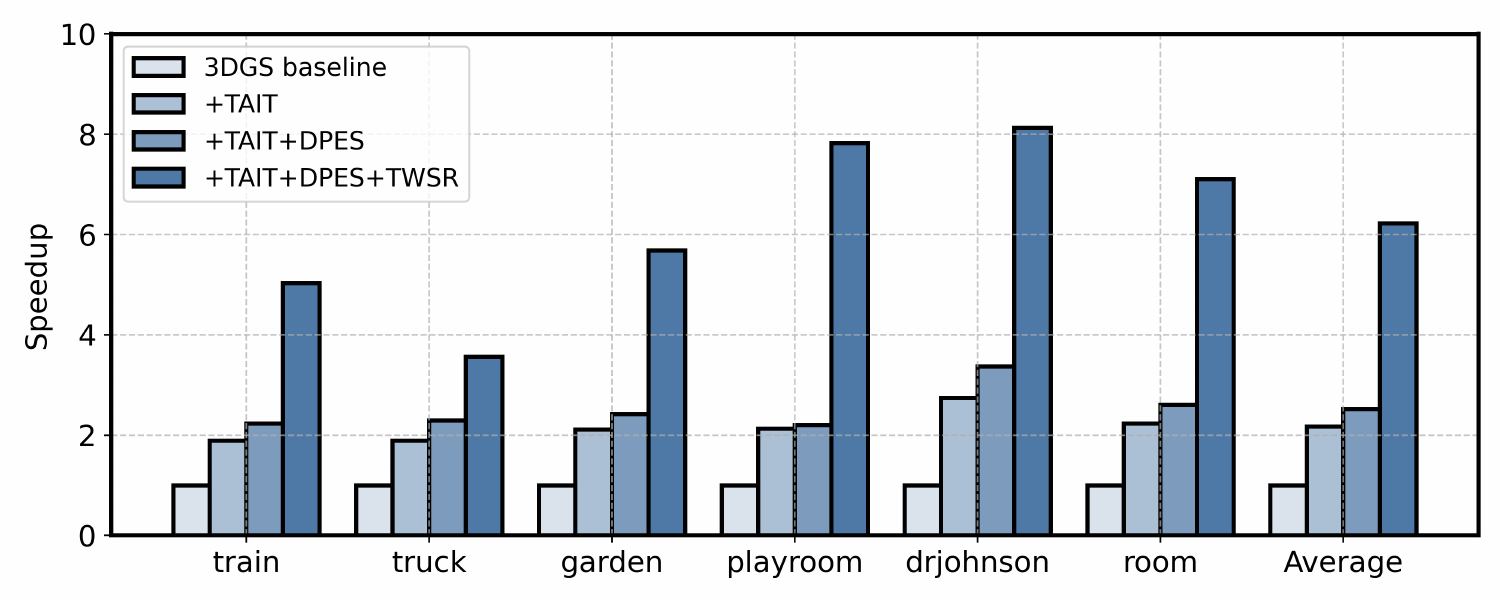}
        \label{GPU ablation study}
    }
    \caption{(a) Comparison of LS-Gaussian performance against prior works across multiple datasets on Jetson AGX Orin and (b) ablation study with algorithm optimizations on real-world scenes.}
    \label{fig13}
\end{figure}

We evaluate the performance of LS-Gaussian deployed directly on the GPU platform using Jetson AGX Orin, as shown in \fref{GPU performance}. Experimental results demonstrate that LS-Gaussian achieves an average speedup of 5.41× over the GPU baseline, satisfying the 90 FPS real-time rendering requirement on the Deep Blending dataset. Compared with AdR-Gaussian and SeeLe, LS-Gaussian also achieves 1.85× and 1.75× speedups, respectively. The most significant improvements are observed in indoor scenes such as playroom and drjohnson from the Deep Blending dataset. This can be attributed to the flattened structures and uniform color patterns typical of indoor environments like floors and walls, which offer higher view consistency and are more amenable to sparse rendering.

To verify this observation, we conduct an ablation study across six real-world scenes, as shown in \fref{GPU ablation study}. We progressively integrate our algorithmic optimization techniques into the original 3DGS pipeline. The introduction of Tile Warping-based Sparse Rendering (TWSR) results in 1.56-2.35× speedups in the three outdoor scenes (train, truck, garden), while achieving 2.41-3.55× speedups in indoor scenes, confirming that the characteristics of indoor environments are more conducive to sparse rendering. The Two-stage Accurate Intersection Test (TAIT) provides an approximate 2× speedups across all scenes, highlighting its general  effectiveness in reducing spurious Gaussian-tile pairs. Notably, beyond enabling accurate tile-level load estimation, Depth Prediction of Early Stopping (DPES) offers a modest speedup by saving preprocessing and sorting overhead for the next frame through depth-based culling.

\subsection{Performance with Hardware Support}
\begin{figure}[!tb]
    \centering
    \includegraphics[width=0.95\linewidth]{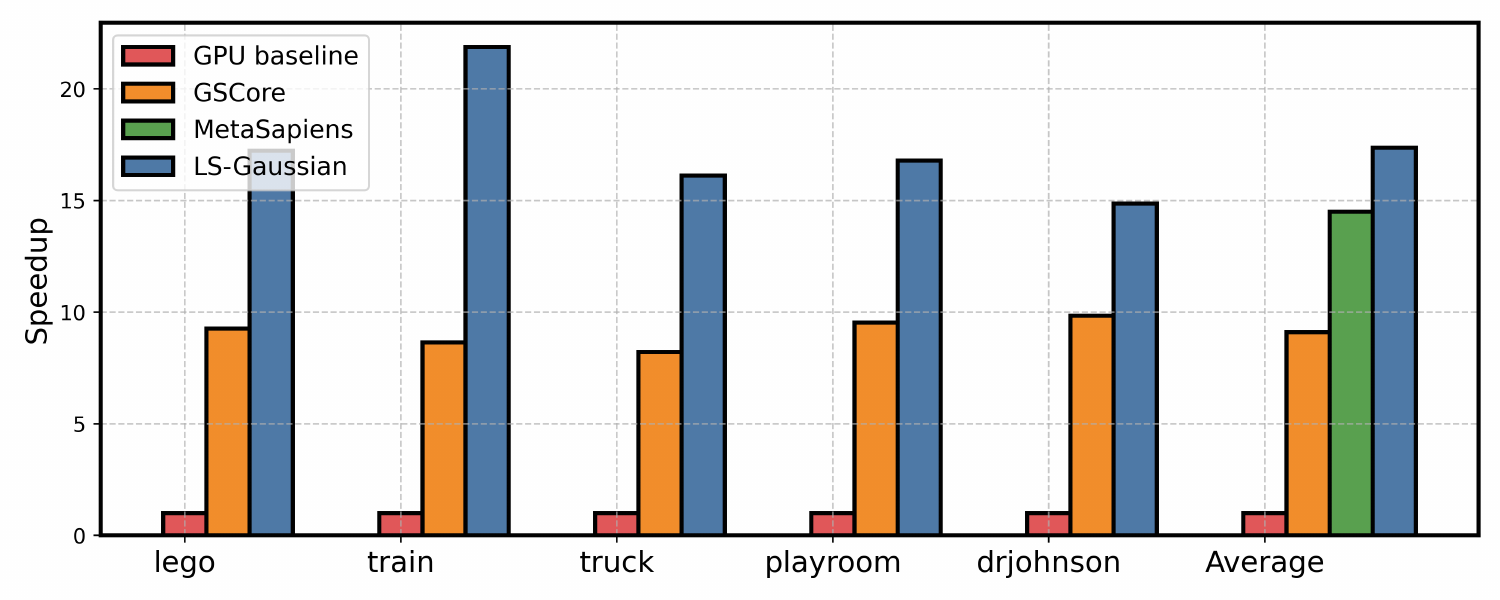}
    \caption{Speedup comparison between our hardware and prior architectures.}
    \label{Hardware speedup}
\end{figure}

\begin{figure}[!tb]
    \centering    
    \subfloat[][]{
	\includegraphics[width=0.46\linewidth]{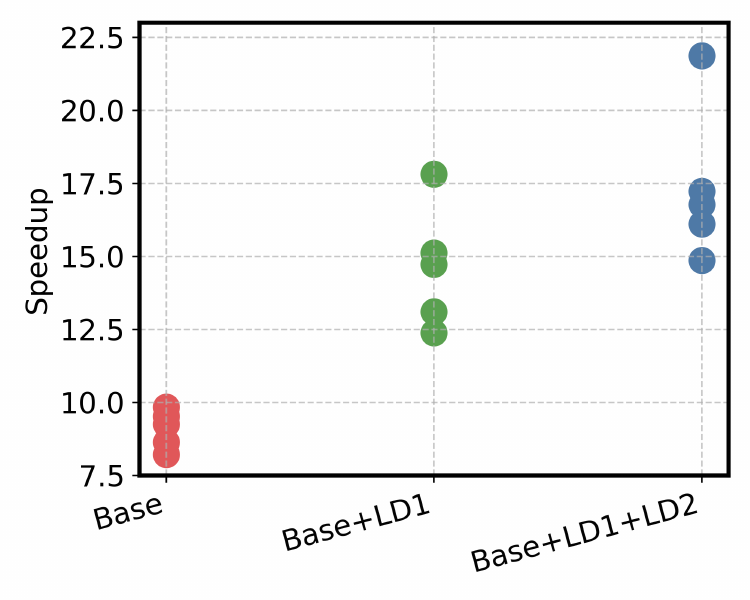}
        \label{Hardware ablation study}
    }
    \vspace{0.01\textwidth}
    \subfloat[][]{
	\includegraphics[width=0.46\linewidth]{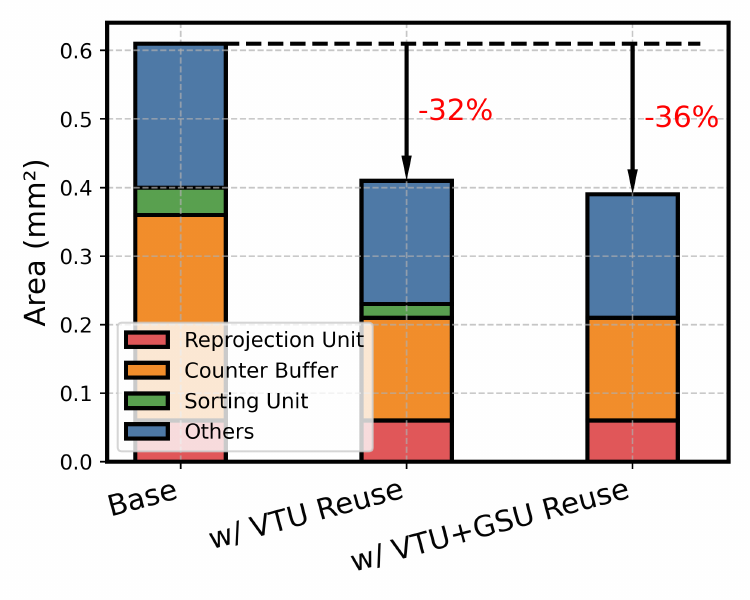}
        \label{Area}
    }
    \caption{(a) Ablation study on the accelerator speedup with inter-block load (LD1) and intra-block load distribution (LD2), and (b) area savings of the augmented units when adopting hardware reuse.}
    \label{fig15}
\end{figure}

We compare our LS-Gaussian with dedicated architectural design against two recent 3DGS accelerators, GSCore and MetaSapiens, in terms of speedup over the GPU baseline. To ensure fairness, we normalize the performance of GSCore and MetaSapiens to the same 1.45 mm\textsuperscript{2} area as LS-Gaussian using the Speedup-Area Curve reported by MetaSapiens. Since MetaSapiens does not report  specific speedups in each scene, we conduct our experiments on scenes from Synthetic-NeRF, Tanks \& Temples and Deep Blending, which are the same as GSCore and MetaSapiens, only reporting the average speedup of MetaSapiens. As shown in \fref{Hardware speedup}, LS-Gaussian achieves an average speedup of 17.3×, outperforming GSCore at 9.1× and MetaSapiens at 14.5×. We ablate the contribution of our base architecture and load distribution strategy from inter- to intra-block level, as illustrated in \fref{Hardware ablation study}, this improvement stems from our specialized hardware design for each processing stage and the balanced load distribution. Besides, we analyze how the hardware reuse strategy in the Load Distribution Unit (LDU) helps reduce the overall accelerator area. As shown in \fref{Area}, we report the area overhead of the augmented modules relative to GSCore. By reusing the counter buffer and comparators from the Viewpoint Transformation Unit (VTU), we achieve a 32\% reduction in additional area cost. Further reuse of Gaussian Sorting Unit (GSU) increases the total savings to 36\%, resulting in only 0.39 mm\textsuperscript{2} of extra area.

As summarized in \tref{tab1}, we further report the average utilization of hardware units across datasets. In contrast to sparse rendering, which favors indoor scenes, balanced load distribution yields greater performance gains in outdoor datasets such as Tanks \& Temples. Unlike indoor environments, outdoor scenes exhibit significant variation in detail across regions, resulting in large workload gaps between high-frequency objects and low-frequency backgrounds, which necessitates carefully designed workload scheduling. 

\begin{table}[!tb]
\begin{center}
\caption{Rasterization core utilization(\%) comparison between original architecture and our LS-Gaussian on Synthetic-NeRF(Synthetic), Tanks \& Temples (T\&T), DeepBlending  (DB), and Mip-NeRF 360  (Mip) datasets.}
\centering
\setlength{\tabcolsep}{8pt} 
\renewcommand{\arraystretch}{1.2}
\resizebox{\linewidth}{!}{
\begin{tabular}{c|ccccc}
\toprule
Method & Synthetic & T\&T & DB & Mip & Average\\
\midrule
Original & 45.6 & 43.1 & 49.5 & 67.9  & 51.5 \\
\hline
LS-Gaussian & 88.5 & 89.8 & 78.0 & 98.2  & 88.6 \\
\bottomrule
\end{tabular}
}
\label{tab1}
\end{center}
\end{table}


\section{Conclusion}
In this paper, we present LS-Gaussian, a lightweight and streaming framework designed to accelerate 3D Gaussian Splatting on resource-constrained platforms. Through a systematic analysis of bottlenecks across the entire 3DGS pipeline, LS-Gaussian introduces a suite of algorithmic optimizations and architectural designs to eliminate redundant computation and mitigate hardware stalls. Extensive evaluations on both synthetic and real-world datasets demonstrate that LS-Gaussian delivers an average speedup of 5.41$\times$ over the edge GPU baseline, and up to 17.3$\times$ acceleration with dedicated hardware support, while maintaining high rendering quality. 


\bibliographystyle{IEEEtran}
\bibliography{reference}

\begin{thebibliography}{10}
\providecommand{\url}[1]{#1}
\csname url@samestyle\endcsname
\providecommand{\newblock}{\relax}
\providecommand{\bibinfo}[2]{#2}
\providecommand{\BIBentrySTDinterwordspacing}{\spaceskip=0pt\relax}
\providecommand{\BIBentryALTinterwordstretchfactor}{4}
\providecommand{\BIBentryALTinterwordspacing}{\spaceskip=\fontdimen2\font plus
\BIBentryALTinterwordstretchfactor\fontdimen3\font minus \fontdimen4\font\relax}
\providecommand{\BIBforeignlanguage}[2]{{%
\expandafter\ifx\csname l@#1\endcsname\relax
\typeout{** WARNING: IEEEtran.bst: No hyphenation pattern has been}%
\typeout{** loaded for the language `#1'. Using the pattern for}%
\typeout{** the default language instead.}%
\else
\language=\csname l@#1\endcsname
\fi
#2}}
\providecommand{\BIBdecl}{\relax}
\BIBdecl

\bibitem{katragadda2024nerf}
S.~Katragadda, W.~Lee, Y.~Peng, P.~Geneva, C.~Chen, C.~Guo, M.~Li, and G.~Huang, ``{NeRF-VINS}: {A} real-time neural radiance field map-based visual-inertial navigation system,'' in \emph{Proc. of IEEE International Conference on Robotics and Automation (ICRA)}, 2024, pp. 10\,230--10\,237.

\bibitem{jiang2024vr}
Y.~Jiang, C.~Yu, T.~Xie, X.~Li, Y.~Feng, H.~Wang, M.~Li, H.~Y.~K. Lau, F.~Gao, Y.~Yang, and C.~Jiang, ``{VR-GS:} {A} physical dynamics-aware interactive {G}aussian splatting system in virtual reality,'' in \emph{Proc. of the ACM SIGGRAPH Conference and Exhibition On Computer Graphics and Interactive Techniques (SIGGRAPH)}, 2024, p.~78.

\bibitem{cao2024lightning}
J.~Cao, Z.~Li, N.~Wang, and C.~Ma, ``{Lightning NeRF}: Efficient hybrid scene representation for autonomous driving,'' in \emph{Proc. of IEEE International Conference on Robotics and Automation (ICRA)}, 2024, pp. 16\,803--16\,809.

\bibitem{tonderski2024neurad}
A.~Tonderski, C.~Lindstr{\"{o}}m, G.~Hess, W.~Ljungbergh, L.~Svensson, and C.~Petersson, ``{NeuRAD}: Neural rendering for autonomous driving,'' in \emph{Proc. of the {IEEE/CVF} Conference on Computer Vision and Pattern Recognition (CVPR)}, 2024, pp. 14\,895--14\,904.

\bibitem{qi2024air}
Z.~Qi, S.~Yuan, F.~Liu, H.~Cao, T.~Deng, J.~Yang, and L.~Xie, ``{AIR-Embodied}: An efficient active {3DGS}-based interaction and reconstruction framework with embodied large language model,'' \emph{arXiv preprint arXiv:2409.16019}, 2024.

\bibitem{huang2025enerverse}
S.~Huang, L.~Chen, P.~Zhou, S.~Chen, Z.~Jiang, Y.~Hu, P.~Gao, H.~Li, M.~Yao, and G.~Ren, ``{EnerVerse}: Envisioning embodied future space for robotics manipulation,'' \emph{arXiv preprint arXiv:2501.01895}, 2025.

\bibitem{chen2025splat}
T.~Chen, O.~Shorinwa, W.~Zeng, J.~Bruno, P.~M. Dames, and M.~Schwager, ``{Splat-Nav}: Safe real-time robot navigation in {G}aussian splatting maps,'' \emph{IEEE Transactions on Robotics}, 2025.

\bibitem{zhao2024drivedreamer4d}
G.~Zhao, C.~Ni, X.~Wang, Z.~Zhu, X.~Zhang, Y.~Wang, G.~Huang, X.~Chen, B.~Wang, Y.~Zhang, W.~Mei, and X.~Wang, ``{DriveDreamer4D}: World models are effective data machines for 4d driving scene representation,'' \emph{arXiv preprint arXiv:2410.13571}, 2024.

\bibitem{ni2024recondreamer}
C.~Ni, G.~Zhao, X.~Wang, Z.~Zhu, W.~Qin, G.~Huang, C.~Liu, Y.~Chen, Y.~Wang, X.~Zhang, Y.~Zhan, K.~Zhan, P.~Jia, X.~Lang, X.~Wang, and W.~Mei, ``{ReconDreamer}: Crafting world models for driving scene reconstruction via online restoration,'' \emph{arXiv preprint arXiv:2411.19548}, 2024.

\bibitem{mildenhall2021nerf}
B.~Mildenhall, P.~P. Srinivasan, M.~Tancik, J.~T. Barron, R.~Ramamoorthi, and R.~Ng, ``{NeRF}: Representing scenes as neural radiance fields for view synthesis,'' in \emph{Proc. of European Conference on Computer Vision (ECCV)}, 2020.

\bibitem{li2022rt}
C.~Li, S.~Li, Y.~Zhao, W.~Zhu, and Y.~Lin, ``{RT-NeRF}: Real-time on-device neural radiance fields towards immersive {AR/VR} rendering,'' in \emph{Proc. of {IEEE/ACM} International Conference on Computer-Aided Design (ICCAD)}, 2022, pp. 132:1--132:9.

\bibitem{kerbl20233d}
B.~Kerbl, G.~Kopanas, T.~Leimk{\"{u}}hler, and G.~Drettakis, ``{3D} {G}aussian splatting for real-time radiance field rendering,'' \emph{ACM Transactions on Graphics (TOG)}, vol.~42, no.~4, pp. 139:1--139:14, 2023.

\bibitem{chen2024survey}
G.~Chen and W.~Wang, ``A survey on {3D} {G}aussian splatting,'' \emph{arXiv preprint arXiv:2401.03890}, 2024.

\bibitem{wu2024recent}
T.~Wu, Y.~Yuan, L.~Zhang, J.~Yang, Y.~Cao, L.~Yan, and L.~Gao, ``Recent advances in {3D} {G}aussian splatting,'' \emph{Computational Visual Media}, vol.~10, no.~4, pp. 613--642, 2024.

\bibitem{li2023instant}
S.~Li, C.~Li, W.~Zhu, B.~T. Yu, Y.~K. Zhao, C.~Wan, H.~You, H.~Shi, and Y.~C. Lin, ``{Instant-3D}: Instant neural radiance field training towards on-device {AR/VR} {3D} reconstruction,'' in \emph{Proc. of Annual International Symposium on Computer Architecture (ISCA)}, 2023, pp. 6:1--6:13.

\bibitem{feng2024cicero}
Y.~Feng, Z.~Liu, J.~Leng, M.~Guo, and Y.~Zhu, ``Cicero: Addressing algorithmic and architectural bottlenecks in neural rendering by radiance warping and memory optimizations,'' in \emph{Proc. of Annual International Symposium on Computer Architecture (ISCA)}, 2024, pp. 1293--1308.

\bibitem{li2024fusion}
S.~Li, Y.~Zhao, C.~Li, B.~Guo, J.~Zhang, W.~Zhu, Z.~Ye, C.~Wan, and Y.~C. Lin, ``{Fusion-3D}: Integrated acceleration for instant {3D} reconstruction and real-time rendering,'' in \emph{Proc. of International Symposium on Microarchitecture (MICRO)}, 2024, pp. 78--91.

\bibitem{liu2024citygaussian}
Y.~Liu, C.~Luo, L.~Fan, N.~Wang, J.~Peng, and Z.~Zhang, ``{CityGaussian}: Real-time high-quality large-scale scene rendering with {G}aussians,'' in \emph{Proc. of European Conference on Computer Vision (ECCV)}, 2024.

\bibitem{fan2024momentum}
J.~Fan, W.~Li, Y.~Han, and Y.~Tang, ``{Momentum-GS}: Momentum {G}aussian self-distillation for high-quality large scene reconstruction,'' \emph{arXiv preprint arXiv:2412.04887}, 2024.

\bibitem{lin2025metasapiens}
W.~Lin, Y.~Feng, and Y.~Zhu, ``{MetaSapiens}: Real-time neural rendering with efficiency-aware pruning and accelerated foveated rendering,'' in \emph{Proc. of International Conference on Architectural Support for Programming Languages and Operating Systems (ASPLOS)}, 2025.

\bibitem{navaneet2024compgs}
K.~L. Navaneet, K.~P. Meibodi, S.~A. Koohpayegani, and H.~Pirsiavash, ``{CompGS}: Smaller and faster {G}aussian splatting with vector quantization,'' in \emph{Proc. of European Conference on Computer Vision (ECCV)}, 2024.

\bibitem{di2025gode}
F.~Di~Sario, R.~Renzulli, M.~Grangetto, A.~Sugimoto, and E.~Tartaglione, ``{GoDe}: {G}aussians on demand for progressive level of detail and scalable compression,'' \emph{arXiv preprint arXiv:2501.13558}, 2025.

\bibitem{fang2024mini}
G.~Fang and B.~Wang, ``{Mini-Splatting}: Representing scenes with a constrained number of {G}aussians,'' in \emph{Proc. of European Conference on Computer Vision (ECCV)}, 2024.

\bibitem{ye20243d}
Z.~Ye, C.~Wan, C.~Li, J.~Hong, S.~Li, L.~Li, Y.~Zhang, and Y.~C. Lin, ``{3D} {G}aussian rendering can be sparser: Efficient rendering via learned fragment pruning,'' in \emph{Proc. of Neural Information Processing Systems (NeurIPS)}, 2024.

\bibitem{zhang2024gaussianspa}
Y.~Zhang, W.~Jia, W.~Niu, and M.~Yin, ``{GaussianSpa}: An ``optimizing-sparsifying'' simplification framework for compact and high-quality {3D} {G}aussian splatting,'' \emph{arXiv preprint arXiv:2411.06019}, 2024.

\bibitem{lu2024scaffold}
T.~Lu, M.~Yu, L.~Xu, Y.~Xiangli, L.~Wang, D.~Lin, and B.~Dai, ``{Scaffold-GS}: Structured {3D} {G}aussians for view-adaptive rendering,'' in \emph{Proc. of the {IEEE/CVF} Conference on Computer Vision and Pattern Recognition (CVPR)}, 2024, pp. 20\,654--20\,664.

\bibitem{ren2024octree}
K.~Ren, L.~Jiang, T.~Lu, M.~Yu, L.~Xu, Z.~Ni, and B.~Dai, ``{Octree-GS}: Towards consistent real-time rendering with {LOD}-structured {3D} {G}aussians,'' \emph{arXiv preprint arXiv:2403.17898}, 2024.

\bibitem{feng2024flashgs}
G.~Feng, S.~Chen, R.~Fu, Z.~Liao, Y.~Wang, T.~Liu, Z.~Pei, H.~Li, X.~Zhang, and B.~Dai, ``{FlashGS}: Efficient {3D} {G}aussian splatting for large-scale and high-resolution rendering,'' \emph{arXiv preprint arXiv:2408.07967}, 2024.

\bibitem{huang2025seele}
X.~Huang, H.~Zhu, Z.~Liu, W.~Lin, X.~Liu, Z.~He, J.~Leng, M.~Guo, and Y.~Feng, ``{SeeLe}: A unified acceleration framework for real-time {G}aussian splatting,'' \emph{arXiv preprint arXiv:2503.05168}, 2025.

\bibitem{lee2024gscore}
J.~Lee, S.~Lee, J.~Lee, J.~Park, and J.~Sim, ``{GSCore}: Efficient radiance field rendering via architectural support for {3D} {G}aussian splatting,'' in \emph{Proc. of International Conference on Architectural Support for Programming Languages and Operating Systems (ASPLOS)}, 2024.

\bibitem{schonberger2016structure}
J.~L. Schonberger and J.-M. Frahm, ``Structure-from-{M}otion revisited,'' in \emph{Proc. of the {IEEE/CVF} Conference on Computer Vision and Pattern Recognition (CVPR)}, 2016.

\bibitem{feng2024potamoi}
Y.~Feng, W.~Lin, Z.~Liu, J.~Leng, M.~Guo, H.~Zhao, X.~Hou, J.~Zhao, and Y.~Zhu, ``Potamoi: Accelerating neural rendering via a unified streaming architecture,'' \emph{ACM Transactions on Architecture and Code Optimization}, vol.~21, no.~4, pp. 80:1--80:25, 2024.

\bibitem{tao2025gs}
M.~Tao, Y.~Zhou, H.~Xu, Z.~He, Z.~Yang, Y.~Zhang, Z.~Su, L.~Xu, Z.~Ma, R.~Fu \emph{et~al.}, ``{GS-Cache}: A gs-cache inference framework for large-scale {G}aussian splatting models,'' \emph{arXiv preprint arXiv:2502.14938}, 2025.

\bibitem{wang2024adr}
X.~Wang, R.~Yi, and L.~Ma, ``{AdR-Gaussian}: Accelerating {G}aussian splatting with adaptive radius,'' in \emph{Proc. of the ACM SIGGRAPH Conference and Exhibition on Computer Graphics and Interactive Techniques in Asia (SIGGRAPH Asia)}, 2024.

\bibitem{hanson2024speedy}
A.~Hanson, A.~Tu, G.~Lin, V.~Singla, M.~Zwicker, and T.~Goldstein, ``{Speedy-Splat}: Fast {3D} {G}aussian splatting with sparse pixels and sparse primitives,'' \emph{arXiv preprint arXiv:2412.00578}, 2024.

\bibitem{li2025gaurast}
S.~Li, B.~Keller, Y.~C. Lin, and B.~Khailany, ``{GauRast}: Enhancing {GPU} triangle rasterizers to accelerate {3D} {G}aussian splatting,'' \emph{arXiv preprint arXiv:2503.16681}, 2025.

\bibitem{ye2025gaussian}
Z.~Ye, Y.~Fu, J.~Zhang, L.~Li, Y.~Zhang, S.~Li, C.~Wan, C.~Wan, C.~Li, S.~Prathipati \emph{et~al.}, ``{G}aussian blending unit: An edge {GPU} plug-in for real-time {G}aussian-based rendering in {AR/VR},'' \emph{arXiv preprint arXiv:2503.23625}, 2025.

\bibitem{knapitsch2017tanks}
A.~Knapitsch, J.~Park, Q.-Y. Zhou, and V.~Koltun, ``Tanks and temples: Benchmarking large-scale scene reconstruction,'' \emph{ACM Transactions on Graphics (TOG)}, vol.~36, no.~4, pp. 1--13, 2017.

\bibitem{hedman2018deep}
P.~Hedman, J.~Philip, T.~Price, J.-M. Frahm, G.~Drettakis, and G.~Brostow, ``Deep blending for free-viewpoint image-based rendering,'' \emph{ACM Transactions on Graphics (TOG)}, vol.~37, no.~6, pp. 1--15, 2018.

\bibitem{barron2022mip}
J.~T. Barron, B.~Mildenhall, D.~Verbin, P.~P. Srinivasan, and P.~Hedman, ``{Mip-NeRF} 360: Unbounded anti-aliased neural radiance fields,'' in \emph{Proc. of the {IEEE/CVF} Conference on Computer Vision and Pattern Recognition (CVPR)}, 2022.

\end{thebibliography}
\vspace{12pt}

\end{document}